**Exciton properties: learning from a decade of measurements on halide perovskites and transition metal dichalcogenides**

*Kameron R. Hansen\*, John S. Colton\*, Luisa Whittaker-Brooks\**

K.R. Hansen, L. Whittaker-Brooks
Department of Chemistry, University of Utah, Salt Lake City, Utah 84112, United States
Email: kameron.hansen@utah.edu; luisa.whittaker@utah.edu
J.S. Colton
Department of Physics and Astronomy, Brigham Young University, Provo, Utah 84602, USA
Email: john_colton@byu.edu

Keywords: exciton, exciton binding energy, band gap energy, perovskite, transition metal dichalcogenide, organic semiconductor, modulation spectroscopy

The exciton binding energy ($E_b$) is a key parameter that governs the physics of many optoelectronic devices. At their best, trustworthy and precise measurements of $E_b$ challenge theoreticians to refine models, are a driving force in advancing the understanding a material system, and lead to efficient device design. At their worst, inaccurate $E_b$ measurements lead theoreticians astray, sew confusion within the research community, and hinder device improvements by leading to poor designs. This review article seeks to highlight the pros and cons of different measurement techniques used to determine $E_b$, namely, temperature-dependent photoluminescence, resolving Rydberg states, electroabsorption, magnetoabsorption, scanning tunneling spectroscopy, and fitting the optical absorption. Due to numerous conflicting $E_b$ values reported for halide perovskites (HP) and transition metal dichalcogenides (TMDC) monolayers, an emphasis is placed on highlighting these measurements in attempt to reconcile the variance between different measurement techniques. By considering the published data en masse, we argue the experiments with the clearest indicators are in agreement on the following values: ~350 - 450 meV for TMDC monolayers between $SiO_2$ and vacuum, ~150 - 200 meV for hBN-encapsulated TMDC monolayers, ~200 - 300 meV for common lead-iodide 2D HPs, and ~10 meV for methylammonium lead iodide.

## 1. Introduction

The two most intensely researched semiconductors of this decade, halide perovskites (HP) and transition metal dichalcogenides (TMDC), share a commonality in that the nature and the role of *excitons* has been a central, on-going question throughout the research sequence. [1, 2] Excitons are electron-hole quasiparticles that form when an electron in the conduction band and a hole in the valence band bind together through their mutual Coulomb attraction. When exciton populations are present, they are responsible for light absorption, light emission, and charge transport; as such, the exciton's properties (mass, radius, binding energy, lifetime, etc.) are central in determining the optical and electronic properties of the host material. Thus, it is of no surprise that accurately measuring the exciton properties in HPs and TMDCs has been the focus of significant research efforts as the community seeks to turn these materials into efficient optoelectronic devices.

Of the exciton's properties, the binding energy ($E_b$) is the most important as it determines the extent to which exciton populations are present within the material— the ratio of the free electron ($N_e$) to exciton ($N_{ex}$) population is given by $N_e/N_{ex} \propto \exp(-E_b/k_BT)$ where $k_BT$ is the thermal energy. Additionally, $E_b$ is also a strong correlate with the remainder of the exciton's other properties, such as radius, oscillator strength, lifetime, etc. [3, 4] However, the value of $E_b$ is both difficult to measure and difficult to accurately predict from first principles calculations. On the experimental side, determining $E_b$ is challenging because the values are small (typically about 0 - 500 meV) and because the measurements often rely on accurate interpretation of band-edge spectral features that are challenging to resolve and have somewhat ambiguous physical origins. Unfortunately, this difficulty in accurately interpreting band-edge spectral features has led to situations where the literature has contradictory information, and it is often difficult to know which $E_b$ measurements are most trustworthy until an appreciable number of studies are published and the data is considered in its totality.

The variance in reported $E_b$ values is particularly large for HPs and TMDCs, even in their two-dimensional forms where confinement effects enhance $E_b$ by an order of magnitude and should make measurements of $E_b$ more feasible. In many cases, the experimental indicators leading to the $E_b$ measurement are in near-perfect agreement study-to-study but are interpreted differently due to ambiguity in their physical origins. [5]

Furthermore, it has been shown different measurement methods yield inconsistent $E_b$ values, even when performed on the exact same sample. [6] Thus, the large the variance in reported $E_b$ is a primarily a consequence of measurement uncertainty rather than discrepancies between samples.

To make sense of these conflicting measurements, we compile extensive lists of $E_b$ measurements for many important excitonic semiconductors—with a particular emphasis on HPs and TMDCs—and discuss the strengths and weaknesses of the most commonly utilized techniques to measure $E_b$. We find the techniques using the clearest indicators yield $E_b$ values that are in agreement for a given material. However, many commonly-utilized techniques such as temperature-dependent photoluminescence and absorption fitting are less reliable and we argue these techniques should not be trusted unless they are complimented by additional experiments.

This article is organized as follows: first, a background in excitons and their optical features is provided. Next, we summarize $E_b$ measurements for several classes of materials, (i.e. III-V semiconductors, TMDCs, HPs, and organic semiconductors) and discuss the unique challenges associated with measuring $E_b$ for each class of material. The bulk of the review (**Sections 4 – 8**) is organized by measurement technique. Six measurement techniques are introduced and discussed, namely temperature-dependent photoluminescence (PL), resolving Rydberg states, electroabsorption (EA), magneto-optical spectroscopy, scanning tunneling spectroscopy (STS), and fitting the optical absorption. By considering $E_b$ measurements from both the "material property" and "experimental technique" perspectives, a wholistic picture emerges that reveals which measurements have been the most trustworthy for materials of the past and present, and hopefully this picture will also aid in guiding $E_b$ measurements of future materials.

2. **Background**

Excitons are divided into Frenkel and Wannier types, where the criterion dividing the two types is the exciton's radius ($r$) which roughly can be thought of as the average separation distance between electron and hole. Frenkel excitons are localized ($r < a$, where $a$ is the unit cell length) whereas Wannier excitons have a greater spatial extent ($r > a$). Typically, Frenkel excitons also have larger binding energies ($E_b > 100$ meV) and move throughout the material via thermal-activated hopping whereas Wannier move via coherent band transport and have lower binding energies ($E_b < 100$ meV); [7] however, Wannier excitons with binding energies as large as ~500 meV been recently been realized in low-dimensional HPs [1] and TMDC monolayers. [2] The existence of Wannier excitons with such a large binding energy in relatively simple structures (natural-forming crystals in the case of 2D HPs) certainly would have come as a shock to condensed matter physicists of the past, and the community is ardently researching these excitons from both fundamental and applied perspectives. Beyond Frenkel and Wannier there is also a third category worthy of mention, that is a charge-transfer exciton, which describes an electron and hole bound together across a material interface. In this article, we are primarily concerned with Wannier excitons.

As with any two-particle system, the Wannier exciton's motion consists of a *center of mass* motion and *relative* motion (see **Figure 1a**). The exciton's center of mass motion behaves like a free particle with mass $M = m_e + m_h$, where $m_e$ ($m_h$) is the electron (hole) effective mass, and the exciton is free to drift throughout the crystal. Unlike the free electron, however, the exciton has neutral charge, integer spin, and obeys Bose-Einstein statistics. The Coulomb-bound state formed between the electron and hole is mathematically analogous to the hydrogen atom. [8] In fact, within the effective mass approximation, the analogy is formally exact. As such, the exciton states are quantized with principal quantum number $n = 1, 2, 3$, etc. and orbital angular momentum $0, \hbar, 2\hbar$, etc. The energy required to create an exciton in an intrinsic semiconductor is the energy needed to excite an electron from the valence band to conduction band (i.e. $E_g$, the fundamental band gap energy; sometimes also referred to as the free particle gap, the one-electron gap), lowered by the energy of the excitonic bound state. For isotropic effective masses, the energy only depends on the principle quantum number $n$, as per: $E_n = E_g - E_b/n^2$ where $E_b$ is also sometimes also referred to as the Rydberg energy in accordance with the hydrogen atom analogy.

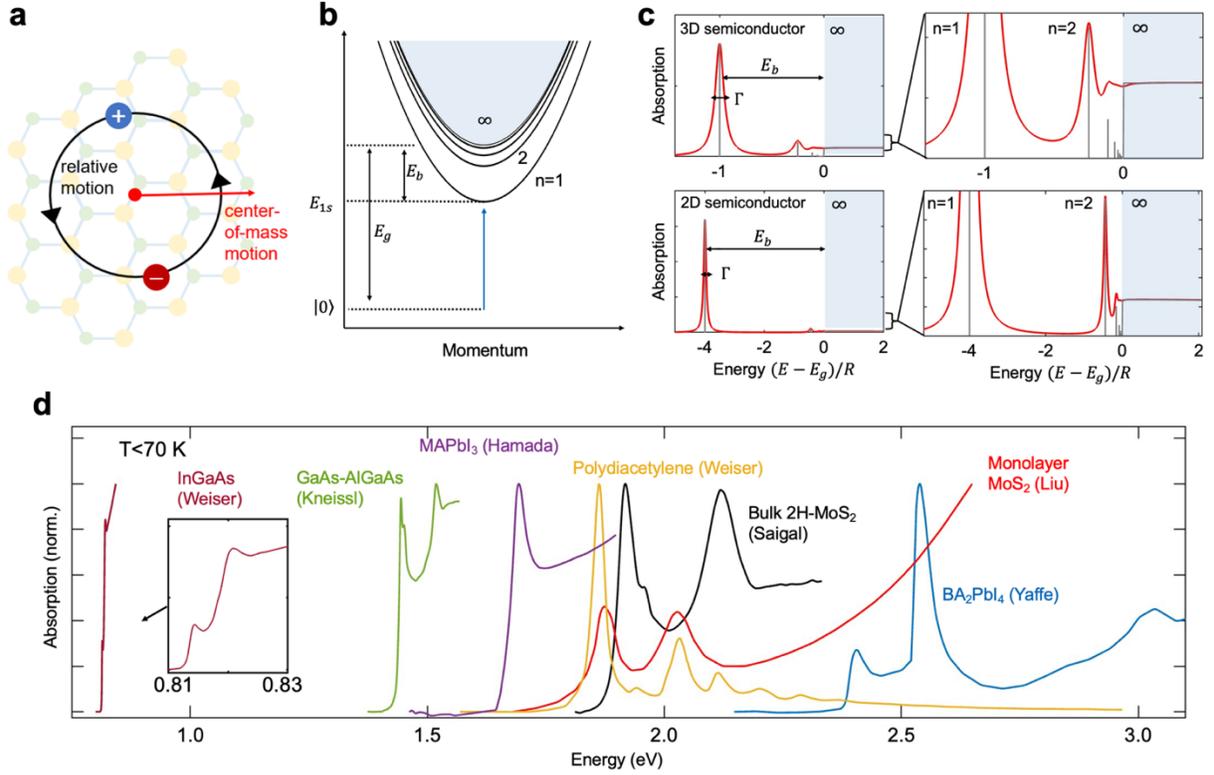

**Figure 1**: An electron-hole pair (exciton) with relative coordinate $r$ that is free to drift throughout the crystal. **(b)** The Rydberg series ($1 < n < \infty$) of a Wannier exciton where the Rydberg energy is represented as $E_b$. The band gap is $E_g = E_\infty$. The vertical axis is energy. **(c)** The theoretical absorption of a Wannier exciton in a three-dimensional semiconductor. Coulomb attraction between the electron and hole increases the interband absorption and results in below-gap absorption peaks with amplitudes $\propto 1/n^3$ where $n$ is the Rydberg number. **(d)** Absorption spectra of commonly studied semiconductors with varying band gap energies, exciton binding energies, and homogenous broadening. All spectra exhibit one or more exciton absorption peaks with excitonic origins. None of the peaks visible to the eye originate from $n > 1$ members of the Rydberg series but instead originate from the 1s exciton state. InGaAs is a three-dimensional III-V semiconductor (spectrum from Weiser et al. [9]); GaAs-AlGaAs is a two-dimensional III-V quantum well structure (spectrum from Kneissl et al. [10]); MAPbI$_3$ is a three-dimensional HP (spectrum from Hamada et al. [11]); polydiacetylene is a one-dimensional organic semiconductor (spectrum from Weiser et al. [9]); bulk 2H-MoS$_2$ is a three-dimensional TMDC (spectrum from Saigal et al. [12]); monolayer MoS$_2$ is a two-dimensional TMDC (spectrum from Liu et al. [13]); BA$_2$PbI$_4$ is a two-dimensional HP (spectrum from Yaffe et al. [14]).

This energy spectrum is shown in **Figure 1b**. Just as the hydrogen atom's Rydberg energy can be written in terms of fundamental constants, so too is the case for the Wannier exciton, but with additional dependence on the material's dielectric constant $\varepsilon_r$ and the exciton's reduced mass $m^* = (m_e m_h)/(m_e + m_h)$, as follows: [8]

$$E_b = \frac{e^4 m^*}{8\pi \hbar^2 \varepsilon_0^2 \varepsilon_r^2} = R_H \left(\frac{m^*}{m_0}\right)\left(\frac{1}{\varepsilon_r^2}\right) \tag{1}$$

where $R_H$ is hydrogen's Rydberg energy (13.6 eV). This Bohr-model expression can be a somewhat accurate approximation for $E_b$ in 3D covalently-bonded semiconductors such as GaAs.

However, this model can fail due to frequency-dependence in $\varepsilon_r$, anisotropic dielectric environments, or when the effective mass approximation breaks down due to confinement effects, and one must seek direct, model-independent measurements of $E_b$.

Experimental techniques to measure $E_b$ center around resolving spectroscopic signatures of the exciton and/or the fundamental band gap. The absorption spectrum of an excitonic semiconductor can be divided into two regions, exciton absorption ($\alpha'_{ex}$) and interband absorption ($\alpha'_{b-b}$):

$$\alpha'(E) = \alpha'_{ex}(E) + \alpha'_{b-b}(E) \tag{2}$$

where $E$ is the photon energy. The exciton component consists of a series of exciton resonances below the band gap, whereas the interband absorption produces a step-like feature, as described by Elliott's formula for three-dimensional semiconductors: [15, 16]

$$\alpha'(E) \propto \sum_{n=1}^{\infty} \frac{2}{n^3} \delta\left(x + \frac{1}{n^2}\right) + \frac{\theta(x)}{1 - e^{-2\pi/\sqrt{x}}} \tag{3}$$

And Elliott's formula for two-dimensional semiconductors: [17]

$$\alpha'(E) \propto \sum_{n=1}^{\infty} \frac{2}{(n-1/2)^3} \delta\left(x + \frac{1}{(n-1/2)^2}\right) + \frac{\theta(x)}{1 + e^{-2\pi/\sqrt{x}}} \tag{4}$$

Here, $\delta$ is the Dirac-delta function, $\theta$ is the Heaviside step function, and $x = (E - E_g)/R$ where $E$ is the photon energy, $E_g$ is the band gap, and $R$ is the exciton's Rydberg energy which is equivalent to the expression for $E_b$ given in Equation (1). For a 3D Wannier exciton, $E_b = R$ whereas $E_b = 4R$ in two dimensions.

These theoretical absorption spectra are plotted in in **Figure 1c** where the Dirac-delta resonances of the exciton's Rydberg series are indicated by gray vertical lines and the interband transition is indicated by a gray step-feature. Of course, the absorption spectrum of a real material will be broadened by exciton's finite lifetime and scattering events within the crystal. Accordingly, a convolution between $\alpha'(E)$ and Lorentzian function $L(E)$ produces a more realistic expression for the absorption spectrum:

$$\alpha(E) = L(E) * \alpha'(E) \tag{5}$$

The solid red line in **Figure 1c** shows **Equation (5)** where the Lorentzian function's full-width-half-maximum (also known as the material's homogenous broadening) has been chosen to have a value of $\Gamma = E_b/10$. Under such conditions when $\Gamma \ll E_b$ and there are no additional spectral features obfuscating the interband absorption, then $E_b$ can sometimes be determined by fitting the linear absorption according to **Equation (5)** where the first and second terms (and oftentimes the Rydberg peaks within the first term) are allowed to vary by scalar factors. However, in reality, these conditions are rarely attained. In fact, we are only aware of a few instances where $E_b$ was accurately measured by fitting to the absorption spectrum.[18-20] Rather, for all the semiconductors discussed in this review, either $\Gamma \sim E_b$ or the presence of the additional band-edge spectral features prevents clear resolution of the interband transition. For most materials, it more feasible to resolve higher members of the exciton's Rydberg series (2s, 3s, etc) than it is to resolve the interband transition in the absorption spectrum.

In order to measure $E_b$, two or more energy levels within **Equation (5)** must be resolved, for example the $n = 1$ exciton and the band gap, or the $n = 1$ and $n = 2$ excitons. The $n = 1$ energy level corresponding to the $1s$ exciton absorption peak is the easiest to resolve, as this feature typically dominates the absorption spectrum. The challenge arises with the second energy level. If the band gap can be resolved, then the binding energy is straightforwardly computed as $E_b = E_g - E_{1s}$. However, if instead higher members of the excitonic Rydberg series are resolved, then one obtains only a lower-bound for $E_b$, e.g. $E_b \geq E_{2s} - E_{1s}$ and a model (the hydrogenic Bohr model or other mode complicated one), must be assumed to provide an estimate of $E_b$. The first situation is to be preferred as the result is then model-independent, but, as is discussed below, obtaining a precise measurement of the band gap is often non-trivial.

To illustrate the deviation of real materials' absorption spectra from that of an ideal Wannier exciton, we have plotted the low-temperature (T < 70 K) absorption spectra of some common excitonic semiconductors in **Figure 1d**. Each spectrum has a large peak corresponding to the exciton's $1s$ energy level. With exception of MAPbI3, each spectrum also exhibits one or more additional peaks that are excitonic in origin. Incidentally however, none of the features visible to the eye in **Figure 1d** correspond to higher members ($n > 1$) of the Rydberg series; instead the additional peaks originate from a variety of other sources, for example excitons in a secondary crystalline phase, vibronic satellites of the $1s$ state, or exciton transitions involving different charge carriers (e.g. heavy vs light holes) or a secondary band gap, i.e. a "multi-valley" absorption peak. These additional peaks overlap with the interband transition, preventing the determination of $E_g$ from the optical absorption. In the case of MAPbI3, the exact energy of interband transition cannot be resolved due to the large homogenous broadening. Thus, more sophisticated forms of spectroscopy are required to measure $E_b$ for each of these technologically-relevant materials.

### 3. Material categories

Generally speaking, the difficulty in measuring $E_b$ for a particular material is determined by the ratio of $E_b/\Gamma$. Materials with larger $E_b$ values are easier to measure because the band-edge absorption features have greater spectroscopic separation. However, this advantage can only be leveraged when the material's broadening $\Gamma$ is low enough for the features to be resolved— in optical absorption experiments, the spectral uncertainty of a given feature is dominated by broadening mechanisms intrinsic to the sample rather than instrumental sources. As such, lowering $\Gamma$ by use of high-quality samples and by cooling to liquid-nitrogen or liquid-helium temperatures is a prerequisite for most of the experimental techniques discussed herein.

If $E_b/\Gamma$ is sufficiently large, then $n > 1$ Rydberg members or the interband transition may by identifiable via traditional spectroscopies (reflection, transmission, absorption, PL, etc.), yielding the "second energy" beyond $E_{1s}$ that is the central challenge in measuring $E_b$. However, categorization of materials based off of $E_b/\Gamma$ alone oversimplifies the situation. For example, in organic semiconductors and TMDC monolayers, $E_b/\Gamma$ is large ($\sim 5-10$), but due to weak interband oscillator strength and the overlap of neighboring band-edge spectral features, the aforementioned traditional spectroscopies typically fail to measure $E_b$. More advanced spectroscopies can at times be used to measure $E_b$, as discussed below; however, these are not without their own challenges. For example, in magneto-optical spectroscopy, the quality of measurement depends more on the exciton's reduced mass $m^*$ more than the ratio of $E_b/\Gamma$. And there are details of sample geometry and synthesis that must be considered as well. For instance, EA spectroscopy might be useful for characterizing exciton properties in TMDC monolayers, but the practical challenge of placing an exfoliated atomic sheet in an electroabsorption device while maintaining low dopant levels has thus far prevented this technique from being widely used on TMDC monolayers.

In short, it is not possible to neatly categorize materials along the criteria of a single parameter or a few parameters in order to decide the optimal technique for measuring $E_b$. Instead, we consider four common classes of semiconductors that span a wide range of $E_b$ and $\Gamma$, namely, III-V semiconductors (low-$E_b$, low-$\Gamma$), TMDCs (moderate-$E_b$, moderate-$\Gamma$), HPs (moderate-$E_b$, high-$\Gamma$), and organic semiconductors (high-$E_b$, high-$\Gamma$). A brief overview of exciton characteristics and a history of $E_b$ measurements for each material is provided in the following sections. These four classes of semiconductors were selected on the criteria of, one, presence of excitons and, two, prevalence in the literature. Roughly speaking, these are the direct-gap semiconductors for which the most $E_b$ measurements have been attempted.

Although this article primarily features studies on direct-gap semiconductors, the $E_b$ measurement strategies for indirect-gap materials are similar. Typically for an indirect gap material, one measures $E_b$ at the lowest direct-edge according to the methods specified herein. [21, 22] Afterwards, $E_b$ at the indirect edge can be estimated by the difference in the reduced effective mass. For some high-quality samples such as crystalline-Si, phonon-assisted exciton absorption can be resolved at the indirect edge using wavelength derivative spectroscopy measurements. [23, 24] And in these cases, $E_b$ can be measured either by fitting the absorption edge or resolving a series of Rydberg peaks similar to direct-gap methods. [25]

**3.1 III-V semiconductors**

The III-V compound semiconductors are formed by combining elements from group III (Al, Ga, In) and group V (N, P, As, Sb) in the periodic table. These elements tend to form wurtzite or zinc blende crystal structures (see **Figure 2a**) with (typically direct) band gaps at room temperature ranging from 0.17 eV (InSb) to 6.0 eV (AlN). In terms of $E_b$ measurements, III-V semiconductors are outliers because both $E_b$ and Γ are an order-of-magnitude lower in comparison to the other materials discussed in this article. The covalent nature of the III-V bond results in large dielectric values at high-frequencies ($\varepsilon_\infty \sim 15$), [26] and $E_b$ is correspondingly quite low, around $1 - 25$ meV (see **Table 1**). Fortunately, in spite of this the exciton properties can still be accurately characterized due to the extremely low homogenous broadening parameter Γ, which can be as low as 0.5 meV in extremely high-purity crystals grown with molecular beam epitaxy. [18] The band-edge absorption of one such high-purity GaAs wafer is shown in **Figure 2a**, where $\Gamma = 0.5$ meV and $E_b = 4.2 \pm 0.2$ meV, and the 1s and 2s exciton absorption peaks are resolved. The small amount of electron-phonon scattering in III-V semiconductors, which is also a result of their covalent nature, is another important aspect that allows for such sharp absorption features.

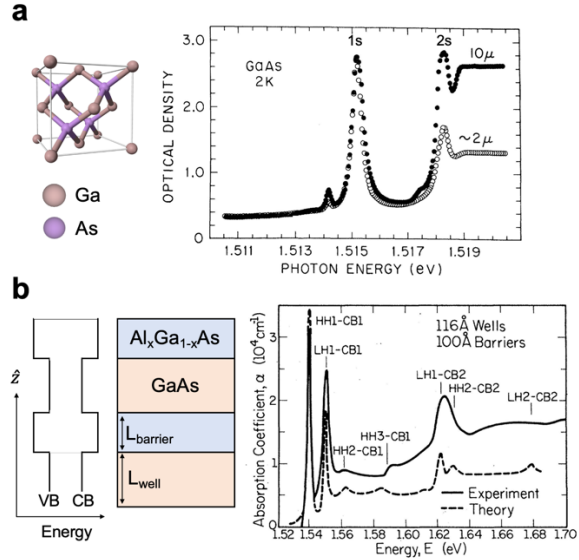

**Figure 2**: **(a)** A depiction of GaAs which forms in the zinc blende crystal structure (left) and the band-edge absorption of a high-purity GaAs wafer at 2K (right). Adapted from ref. [18] with permission from the American Physical Society, Copyright 1972. **(b)** A common III-V quantum well structure with alternating high-$E_g$ (Al$_x$Ga$_{1-x}$As) and low-$E_g$ (GaAs) layers (left) and the corresponding band-edge absorption of the GaAs layer (right). Adapted from ref. [27] with permission from the American Physical Society, Copyright 1985.

Advancements in molecular beam epitaxy throughout the 1970s and 1980s allowed researchers to create a diverse range of III-V heterostructures, such as the quantum well structure displayed in **Figure 2b** with alternating Al$_x$Ga$_{1-x}$As and GaAs layers. In these structures, the high-$E_g$ layer acts as an energy barrier and confines the electronic states within the low-$E_g$ well. Unlike low-dimensional HP and TMDC structures, there is only a slight mismatch in dielectric constants at the barrier-well junction and therefore image-charge effects are only minor. [28] Still, the quantum confinement is sufficient to increase $E_b$ by a factor of ~5, allowing for stable excitons at room temperature. The band-edge absorption for a representative GaAs/Al$_x$Ga$_{1-x}$As quantum well is shown in **Figure 2b**, where the features are blueshifted by ~ 25 meV with respect to the GaAs wafer absorption in **Figure 2a** due to confinement effects. Here, the well length ($L_w$) and barrier length ($L_b$) are 11.6 and 10 nm respectively. For comparison, the band edge of a sample with $L_w = L_b = 5$ nm is blueshifted by ~ 140 meV. [29]

The multiple exciton peaks in **Figure 2b** are not assigned to the 1$s$ and 2$s$ states, but rather to distinct 1s excitons associated with heavy-hole and light-hole transitions. Most III-V compounds have two *nearly* degenerate valance band maximums with differing effective masses which leads to two distinct interband transitions with only ~ 30 meV separation and two distinct $E_b$ values according to **Equation (1)**; the heavy hole (HH) has a greater $E_b$ than the light hole (LH). [30] Naturally, the presence of multiple band-edge features increases the difficulty in resolving and assigning peaks, and therefore makes $E_b$ measurements more challenging. However, due to the extremely low Γ values that can be obtained in high-purity samples, excellent $E_b$ measurements can still be obtained by resolving $n > 1$ or using magnetic-field induced shifts to the exciton's 1s energy peak. Multiple studies have used such magneto-optical probes to measure $E_b$ as a function of well-width and the results are in good agreement with one another. [30-32]

**Table 1**: Exciton binding energy measurements in selected III-V semiconductors

| Material | Sample details | Technique | $E_b$ (meV) | Reference |
|---|---|---|---|---|
| | | Bulk | | |
| GaAs | | Trans., Rydberg | 4.2 ± 0.2 | Sell [18] |
| GaAs | | Trans., Rydberg | 4.1 | Hill [20] |
| GaAs | | EA | 3.8 | Ziffer et al. [33] |
| GaAs | | Magnetoreflection | 4.2 ± 0.3 | Nam et al. [34] |
| GaP | | PLE | 21 ± 2 | Kopylov et al. [35] |
| InAs | | Magnetoabsorption | 1 | Tang et al. [36] |
| InP | | Absorption | 4 | Turner et al. [37] |
| GaAs | | PL | 4.7 ± 0.4 | Gilleo et al. [38] |
| GaN | | Temp-dependent PL | 26 | Viswanath et al. [39] |
| InSb | | Magnetoabsorption | 0.7 | Pidgeon et al. [40] |
| | | Quantum well | | |
| GaAs/Al$_x$Ga$_{1-x}$As | 4<L<22 nm | PLE, Rydberg | 7<$E_b$<11.5 (LH) | Koteles et al. [41] |
| | | | 6<$E_b$<10.5 (HH) | |
| GaAs/Al$_x$Ga$_{1-x}$As | 4<L<10 nm | Magneto-PLE | 12<$E_b$<17 (HH) | Maan et al. [30] |
| GaAs/Al$_x$Ga$_{1-x}$As | 9<L<12 nm | Magneto-PLE | 9<$E_b$<10 (LH) | Maan et al. [30] |
| GaAs/AlAs | 6<L<25 nm | Magnetoabsorption | 5<$E_b$<14.5 (HH) | Tarucha et al. [31] |
| GaAs/Al$_x$Ga$_{1-x}$As | 8<L<30 nm | Magneto-PL | 5.2<$E_b$<7.4 (HH) | Ossau et al. [32] |
| GaAs/Al$_x$Ga$_{1-x}$As | 1<L<10 nm | PLE | 3.8<$E_b$<9 (HH) | Chomette et al. [42] |
| GaAs/In$_x$Ga$_{1-x}$As | 1.4<L<4.8 nm | PLE | 6.5<$E_b$<9.2 (HH) | Moore et al. [43] |
| GaAs/In$_x$Ga$_{1-x}$As | L = 8 nm | Magnetoabsorption | 6.3 | Hou et al. [44] |

**3.2 Transition-metal dichalcogenides**

Transition-metal dichalcogenides (TMDCs) with general formula MX$_2$ consist of a transition metal (M = Mo or W) bonded to a chalcogenide (X = S, Se, or Te). The bulk crystal structure, shown in **Figure 3a**, is layered in nature because the covalently-bonded MX$_2$ sheets are held together by van der Waals forces along the c-axis. The covalent nature of the M-X bond results in relatively large high-frequency dielectric constants ($\varepsilon_\infty \approx 10$),[45-47] thus one would expect a low $E_b$ in TMDC systems. However, the CB and VB dispersion at the direct transition where excitons form is relatively flat, resulting in large effective masses and correspondingly large $E_b$ values ($\sim 50 - 100$ meV for bulk TMDCs, see **Table 2**).

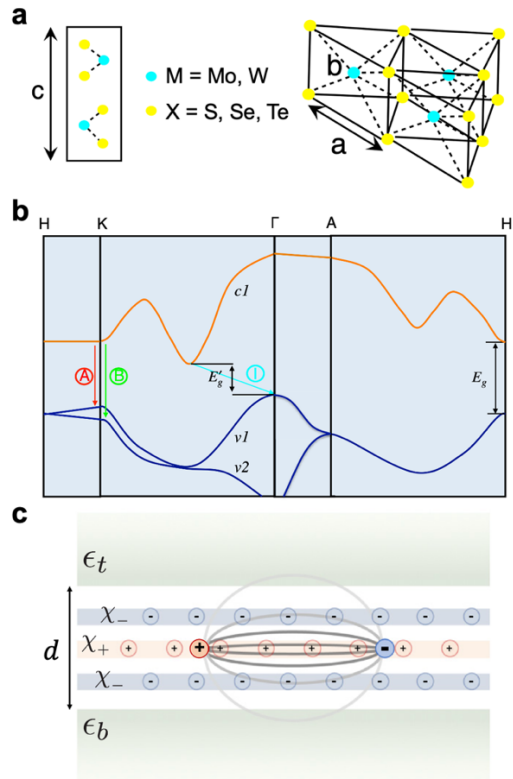

**Figure 3**: **(a)** TMDC lattice structure with general formula MX$_2$. The M and X sites have an covalently bonded along the a- and b-axes, but are only held together by van der Waals interactions along the c-axis. **(b)** A simplified band structure of bulk MoS$_2$. The band gap lies at the indirect transition labeled 'I'. The direct transitions 'A' and 'B' blueshift less than the 'I' transition in the presence of quantum confinement due to their larger out-of-plane mass and as a result MoS$_2$ transforms to a direct-gap material when exfoliated down to a single layer. **(c)** Depiction of an exciton in a TMDC monolayer where the Coulomb interaction is weakly screened by the MX$_2$ layer and the top and bottom dielectric environments. Panels (a-b) are reprinted from ref. [48] with permission from the American Physical Society, Copyright 2010. Panels (c) is from ref. [49], reprinted with permission from the American Physical Society, Copyright 2018.

A simplified band structure for bulk MoS$_2$, in **Figure 3b**, depicts this flat-band-nature of the direct transitions labeled 'A' and 'B' at the K point in the Brillouin zone. This direct transition results in an absorption edge with exciton features that were studied extensively throughout the 1960s and 1970s using absorption, electroabsorption, and magnetoabsorption spectroscopies. [21, 22, 50-52] However, this direct transition does not mark the band gap for bulk TMDCs since a lower, indirect transition exists. In **Figure 3b**, the indirect transition is labeled 'I' with a gap energy of $E'_g$ which falls between 1.09 eV (MoSe$_2$) and 1.35 eV (WS$_2$). [53] For reference, the direct transition with energy $E_g$ (labeled 'A' in **Figure 3b**) is typically ~ 2 eV. [53] Due to the lower perpendicular electron (hole) mass of the CBM (VBM) participating in the "I" transition, this region of the Brillouin zone is more sensitive to quantum confinement effects. As a result, when the number of layers is reduced, the upshift of $E'_g$ exceeds that of $E_g$ and the material transforms to a direct-gap semiconductor in the monolayer limit. [48]

Monolayer TMDCs (or 2D TMDCs) were first realized in 2010 [48, 54, 55] and since have sparked a wave of research interest in solid-state physics due to their small size, unique exciton properties, and unique spin properties. [56] Exfoliating down to a single layer reduces the TMDC dielectric constant from ~10 to ~5 [45, 46, 57], and TMDC monolayers host Wannier excitons with extremely high binding energies in the range of 150 – 500 meV (see **Table 2**). Unlike III-V quantum wells where the dielectric mismatch is low or 2D HPs where the semiconducting layer is tethered between insulating layers, exfoliated TMDC monolayers are 2D semiconductors that are fully exposed to their surroundings. The dielectric environment on top ($\varepsilon_t$) and below ($\varepsilon_b$) below the monolayer, depicted in **Figure 3c**, strongly modulates the in-plane Coulomb interaction between the electron and hole, according to the Rytova-Keldysh potential [49, 58, 59]:

$$V_{RK}(r) = -\frac{e^2}{r_0}\frac{\pi}{2}\left[H_0\left(\frac{\kappa r}{r_0}\right) - Y_0\left(\frac{\kappa r}{r_0}\right)\right] \quad (6)$$

where $H_0$ and $Y_0$ are the Struve function and the Bessel function of the second kind respectively, $r_0$ is the dielectric screening length, and $\kappa = (\varepsilon_t + \varepsilon_b)/2$. For example, when the monolayer is sandwiched between hexagonal boron nitride (hBN) with $\varepsilon = 4.5$, the Coulomb screening in the barrier is significant and $E_b$ ranges from 167 meV (WS$_2$) to 224 meV (MoS$_2$). [60, 61] However, in the presence of weak Coulomb screening when the monolayer is suspended in vacuum ($\varepsilon_b = \varepsilon_t = 1$) or placed on SiO$_2$ ($\varepsilon_b = 2.1$; $\varepsilon_t = 1$), measurements suggest that $E_b$ is ~ 400 meV [62, 63]. In **Table 2**, various $E_b$ measurements for MoS$_2$, MoSe$_2$, WS$_2$, and WSe$_2$ are tabulated and the information of the surrounding dielectric media is included.

Interestingly, there has been substantial interest in measuring the Rydberg series in TMDC monolayers because the series appears to deviate from the hydrogenic model of $E_n = R/n^2$ in the case of anisotropic screening of Coulomb interaction ($\varepsilon_t \neq \varepsilon_b$). [63, 64] This effect is predicted by the Rytova-Keldysh potential [65, 66] and more sophisticated theories as well. [49] However, when the monolayer is sandwiched between insulating hBN layers ($\varepsilon_t = \varepsilon_b$), the Rydberg series shows no significant deviation from a quasi-2D hydrogenic model. [60, 61] There is also significant interest in stacking TMDC monolayers on top of one another to create heterostructures. If the band alignment and sample quality permits, a photogenerated electron (hole) in one layer will relax to the CBM (VBM) in the neighboring layer while still bound to its pair, creating a so-called interlayer exciton that is somewhat analogous to a charge-transfer exciton. [67-70] While these interlayer excitons have intriguing properties, they have no optical signature, so the best estimates of their binding energies come from STS, PL, and theoretical calculations. [71-73] For the purposes of this article, we are primarily concerned with $E_b$ measurements on *intra*layer excitons in TMDC monolayers.

*Unique challenges in measuring $E_b$:*
- The band gap absorption is buried beneath the 'B' exciton peak, which originates from a spin-orbit-split VB at lower energies (labeled v2 in **Figure 3b**).
- The value of Γ is typically quite large (~50 meV) for TMDC monolayers due to interactions between the monolayer and substrate. However, these effects can be reduced by encapsulating the monolayer in hBN, resulting in Γ as low as ~ 2 meV at 4 K. [64, 74, 75]
- The exciton's reduced effective mass $m^*$ is too large to reach the high-field limit of magnetoabsorption (see **Section 7**).
- The strategy of obtaining $E_b$ by resolving higher members of the Rydberg series has an additional layer of complexity for TMDC monolayers because the correct model must be selected to extrapolate the series to $E_g$. This is only an issue when three or fewer of the $E_n$ states are measured. In the case $n = 4$ and/or $n = 5$ are clearly resolved, then the correct model is experimentally verified, and additionally, $E_g$ is tightly confined by proximity to the $n = 4$ or $n = 5$ state.

*Notable measurements:*

- An unambiguous signal originating from the Rydberg series ($1 \leq n \leq 5$) has been detected in monolayer TMDCs using low-field magneto-PL and magnetoabsorption. Multiple studies have detected this series and the results are in excellent agreement, with $E_b \sim 200$ meV (**Section 6.1**). [60, 61, 76-78]
- He et al. resolved faint, but reproducible oscillations above the A exciton in the reflectance contrast signal of WSe$_2$ monolayer. These were assigned to the exciton's $2s - 5s$ energies since they follow a non-hydrogenic Rydberg ladder and coincide with a broad feature in the two-photon PLE spectrum originating from the $2p$ and $3p$ absorption (see **Section 5.2**). [64] Similar results were obtained by Chernikov et al. on WS$_2$ monolayers. [66]

**Table 2**: Exciton binding energy measurements on TMDC semiconductors

| Material | Sample details | Technique | $E_b$ (meV) | Ref. |
|---|---|---|---|---|
| Bulk | | | | |
| MoS$_2$ | Single crystal flakes (2H) | 2P-Abs | 219 | Xie et al. [79] |
| MoS$_2$ | Single crystal flakes | Abs | 100 ± 7 | Karmakar et al. [80] |
| MoS$_2$ | Single crystal (2H) | Abs, Rydberg | 53 | Evans [50] |
| WS$_2$ | Single crystal flakes (2H) | Refl., Photorefl. | 90 | Jindal et al. [81] |
| WS$_2$ | Single crystal (3R) | Refl., Rydberg | 64 | Beal et al. [82] |
| WSe$_2$ | Single crystal (2H) | Refl., Rydberg | 56 | Beal et al. [82] |
| WSe$_2$ | Single crystal (2H) | Trans., Rydberg | 56 | Mitioglu et al. [83] |
| Monolayer | | | | |
| MoS$_2$ | Exf. on SiO$_2$ | 2P-Abs | 879 | Xie et al. [79] |
| MoS$_2$ | Exf. between hBN | Magneto-PL | 217 | Molas [61] |
| MoS$_2$ | Exf. between hBN | Magnetoabsorption | 221 | Goryca [60] |
| MoS$_2$ | Exf., suspended | PC | ≥570 | Klots et al. [62] |
| MoS$_2$ | Exf. on hBN/fused silica | PLE | 440 ± 80 | Hill et al. [84] |
| MoS$_2$ | CVD on HOPG | STS, PL | 300 | Chiu et al. [85] |
| MoS$_2$ | CVD on HOPG | STS, PL | 420 | Zhang et al. [86] |
| MoS$_2$ | Exf. on fused silica | STS, Refl. | 310 ± 40 | Rigosi et al. [87] |
| MoSe$_2$ | Exf. between hBN | Magneto-PL | 216 | Molas [61] |
| MoSe$_2$ | Exf. between hBN | Magnetoabsorption | 231 | Goryca [60] |
| MoSe$_2$ | CVD on HOPG | STS, PL | 500 | Zhang et al. [88] |
| MoSe$_2$ | MBE on graphene/SiC | STS, PL | 550 | Ugeda et al. [89] |
| MoSe$_2$ | MBE on HOPG | STS, PL | 590 ± 50 | Liu et al. [90] |
| WSe$_2$ | Exf. between hBN | Magneto-PC | 169 | Wang [91] |
| WSe$_2$ | Exf. between hBN | Magneto-PL | 167 | Molas [61] |
| WSe$_2$ | Exf. between hBN | Magneto-PL | 170 | Chen et al. [92] |
| WSe$_2$ | Between SiO$_2$ and hBN | Magnetoabsorption | 201 ± 20 | Steir [77] |
| WSe$_2$ | Between SiO$_2$ and polymer | Magnetoabsorption | 311 ± 40 | Steir [77] |
| WSe$_2$ | Exf. on SiO$_2$ | Magnetoabsorption | 442 ± 60 | Steir [77] |
| WSe$_2$ | Exf. between hBN | Magnetoabsorption | 167 | Goryca [60] |
| WSe$_2$ | Exf. between hBN | Magnetoabsorption | 167 | Stier [76] |
| WSe$_2$ | Exf. on diamond | Mid-IR pump probe | 245 | Pöllmann et al. [93] |
| WSe$_2$ | Exf. on SiO2/Si | PLE, 2P-PLE, SHG | 600 ± 20 | Wang et al. [94] |
| WSe$_2$ | Exf. on SiO2/Si | Refl. | 887 | Hanbicki et al. [95] |
| WSe$_2$ | Exf. on SiO2/Si | Refl., 2P-PLE | 370 | He et al. [64] |
| WSe$_2$ | CVD on HOPG | STS, PL | 500 | Zhang et al. [88] |
| WSe$_2$ | CVD on HOPG | STS, PL | 400 | Chiu et al. [85] |
| WSe$_2$ | MBE on HOPG | STS, PL | 720 ± 70 | Liu et al. [90] |
| WS$_2$ | Exf. on fused silica | 2P-PLE | 700 | Ye et al. [63] |
| WS$_2$ | Exf. on SiO2/Si | 2P-PLE | 710 | Zhu et al. [96] |
| WS$_2$ | Exf. between hBN | Magneto-PL | 174 | Molas [61] |
| WS$_2$ | Exf. between hBN | Magnetoabsorption | 180 | Goryca [60] |
| WS$_2$ | CVD on SiO2 | Magnetoreflection | 410 | Stier et al. [78] |
| WS$_2$ | Exf. on SiO2/Si | Refl. | 320 ± 40 | Chernikov et al. [66] |
| WS$_2$ | Exf. on SiO2/Si | Refl. | 929 | Hanbicki et al. [95] |
| WS$_2$ | Exf. on fused silica | Refl., PLE | 320 | Hill et al. [84] |
| WS$_2$ | Exf. on fused silica | STS, Refl. | 360 | Rigosi et al. [87] |

### 3.3 Halide perovskites

Halide perovskites are a class of solution-processable semiconductors that have risen to prominence over the last decade due largely to their high efficiency in solar cell devices. [97-99] The electronic structure in Halide perovskites originates from the inorganic metal-halide $[BX_6]^{-4}$ octahedra where typically B = $Pb^{2+}$ or $Sn^{2+}$, and X = $I^-$, $Br^-$, or $Cl^-$. In their three-dimensional form with general formula $ABX_3$, the octahedral are corner-sharing (see **Figure 4a**); this creates an open pocket to be filled by an A-site cation (often $Cs^+$ or $MA^+$ where MA = methylammonium) which only weakly interacts with the electronic structure. Due to their relatively large dielectric constants and low carrier masses, 3D HPs host excitons with low binding energies (~7 – 25 meV). [33, 100-102] When the A-site cation is replaced by a larger molecule, however, the octahedra separate into two-dimensional sheets (see **Figure 4b**) and $E_b$ increases an order of magnitude due to confinement effects and an overall decrease in $\varepsilon_r$. Similar to TMDCs which can be processed into bilayer and trilayers, there is a range of inorganic layer thicknesses between the 3D and 2D extrema depicted in **Figure 4a-b** (often referred to as $n = 1, 2, 3$, etc. where $n$ is the number of inorganic [BX$_4$]$^{-2}$ sheets between the spacer cations). However, $E_b$ measurements for the $n > 2$ phases are rare [103, 104] and we focus primarily on the $n = 1$ phase.

For $n = 1$ 2D HPs, the mismatch in band alignment between the low-$E_g$ metal-halide layer and large-$E_g$ organic layer confines the electronic structure to the metal-halide sheet thereby widening gap and decreasing the polarizability. [105, 106] Additionally, there is a mismatch in dielectric constants between the organic barrier layer ($\varepsilon_\infty \sim 2.5$) and the inorganic well layer ($\varepsilon_\infty \sim 5$), which imparts an image charge effect on exciton and further increases its binding energy by a factor of $\sim 2 - 3$. [105-107]

Due to the ionic nature of the metal-halide bond, 2D and 3D HPs have low dielectric values at frequencies above the phonon resonance $\varepsilon_\infty \sim 5$, but large dielectric values when phonons contribute to the polarization, $\varepsilon(f) \sim 20$ for frequencies in the range of 1-1000 GHz. [108, 109] The dielectric constant becomes even larger when the rotation of the dipolar MA cation contributes for frequencies less than 1 GHz, $\varepsilon(f) \sim 50$. [110-112] This leads to a situation where a simple Bohr model prediction of $E_b$ according to **Equation (1)** varies between 0.5 meV and 55 meV for 3D HPs depending which frequency of the dielectric constant is selected. When perovskite solar cell efficiencies dramatically rose in the 2010s, the description of charge carriers as free electrons or excitons was hotly debated. [4, 33, 102, 113] Oftentimes this debate centered around conflicting $E_b$ measurements and the frequency-value of $\varepsilon_r(f)$ that was most relevant in determining $E_b$. [112, 114] In addition to being difficult to estimate from **Equation (1)** using material parameters, $E_b$ is also difficult to measure experimentally in 3D HPs. The ionic nature of the metal-halide bond enhances charge-lattice interactions and leads to frequent scattering of charge carriers. As a consequence, Γ is large (~35 meV at room temperature [33]). Cooling to low temperatures only reduces Γ by about 60-80% because defect scattering is also significant for HPs. Furthermore, 3D HP structures undergo multiple phase transitions (cubic → tetragonal → orthorhombic) moving from 300 K to 0 K, making it challenging to relate $E_b$ measurements in the low temperature phase to the more technologically-relevant room temperature phase.

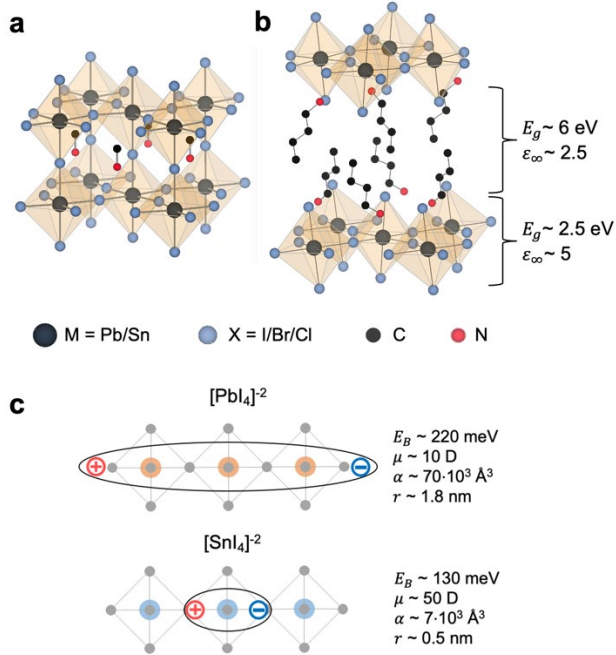

**Figure 4**: **(a)** A three-dimensional HP with general formula $ABX_3$ where A = methylammonium in this case. The hydrogen atoms are removed for clarity. **(b)** A two-dimensional HP with general formular $A_2BX_4$ where A = butylammonium in this case. The large $E_g$ and low $\varepsilon_\infty$ of the organic layer creates quantum and dielectric confinement effects. As a result, excitons in the 2D HPs have binding energies an order of magnitude greater than those in 3D HPs. **(c)** Exciton properties measured for lead and tin-iodide 2D HPs. The disorder associated with the $SnI_4$ framework localizes excitons, thereby reducing their polarizability and increasing their dipole moment. Reprinted from ref. [115] with permission from Wiley-VCH, Copyright 2022.

The first measurements of $E_b$ in 3D HPs came by way of low-field magnetoabsorption, [116, 117] linear absorption fits, [4] and temperature-dependent PL. [118-120] Each of these methods estimated $E_b$ to on the order of 30 – 70 meV. In the following years, more sophisticated techniques such as high-field magnetoabsorption [100, 101] and third-derivative EA fitting [33] revealed more accurate $E_b$ measurements ranging from 7 – 25 meV (see **Table 3**), where the majority of the variance here comes from different chemical compositions and crystalline phases of the $ABX_3$ structure. Exciton binding energy measurements for 2D HPs have followed a similar history. Fitting to the linear absorption [19, 121] and temperature-dependent PL [122, 123] were popular strategies to measure $E_b$ in 2D HPs early on. However, as discussed in **Section 4**, there is mounting evidence that activation energies obtained from Arrhenius-model fitting of temperature-dependent PL don't necessarily correlate with $E_b$ for HPs. In just the past few years, more sophisticated measurements in the form of EA, [115, 124] high-field magnetoabsorption, [103, 125] and photoluminescence excitation (PLE) [126] paint a consistent and convincing picture that $E_b$ is about 250 meV for lead-iodide-based 2D HPs. Other reports have claimed $E_b$ is ~500 meV in these compounds due to faint, near-gap absorption features that may arise from higher members ($n > 1$) of the exciton's Rydberg series (discussed further in **Section 5.1**). [14, 104] In **Table 3**, several binding energy measurements are listed for the most commonly studied lead-iodide perovskite compositions— MAPbI$_3$ represents the 3D HP category, while butylammonium lead iodide (BA$_2$PbI$_4$) and phenethylammonium lead iodide (PEA$_2$PbI$_4$) represent the 2D HP category.

Despite their high binding energies, the excitons in 2D HPs are typically Wannier in nature—recent magnetoabsorption studies by Dyksik et al. [103] and EA studies by Hansen et al. [115, 124] are in agreement that the exciton's radius spans ~3 metal-halide octahedral for lead-iodide perovskites (see **Figure 4c**). However, when lead is replaced with tin, these same experiments detect a substantial decrease in $E_b$ concomitant with a decrease in exciton radius which is contrary to the Wannier model wherein the binding energy and radius are inversely related ($r \propto 1/\sqrt{E_b}$). [115, 125] This deviation from Wannier behavior is a result of charge-lattice interactions which are significant in HPs. The incoherent motion of acoustic phonons introduces disorder into the charge carrier's energy landscape, inducing fluctuations in the band structure [127] and localizing exciton wavefunctions. [115, 128, 129] Meanwhile, the coherent motion in the form of optical phonons couples to excitons, resulting in multiple exciton-polaron states with distinct lattice couplings. [130-132] Similar effects are present in 3D HPs, but charge carriers take the form of free polarons due to the low exciton binding energy. [133, 134] With the hindsight of a decade's worth of research, it's unsurprising that $E_b$ measurements in HPs were initially so challenging. The lattice-induced broadening of optical features and low $E_b$ values make measurements of the exciton's properties extremely tricky in these systems; however, these obstacles motivated advancements in several fields, particularly EA, magnetoabsorption, and density functional theory (DFT), [109] that ultimately that have increased our general understanding of excitons in systems with static and dynamic disorder.

*Unique challenges in measuring $E_b$:*

- The broadening is greater than the binding energy in 3D HPs. For example, for MAPbI3 at room temperature, $\Gamma \sim 40$ meV and $E_b \sim 10$ meV. [33, 113, 135-137]
- Multi-valley absorption leads to exciton peaks above the band gap that bury the interband absorption and complicates fitting the linear absorption to Elliott's formula. [104, 115, 138]
- In 2D HPs, the $1s$ exciton has multiple absorption peaks from distinct lattice couplings that are difficult to interpret and can sometimes overlap with interband transitions. [130, 139-141]

*Notable $E_b$ measurements:*
- Ziffer et al. implemented Aspnes' low-field Franz-Keldysh theory [142] to fit the EA spectrum of MAPbI3 and obtain $E_b = 7.4 \pm 2$ meV in the room-temperature tetragonal phase (**see Section 6.2**). [33]
- Distinct Stark and Franz-Keldysh features have been resolved in the EA spectrum of 2D HPs. Simulations of the EA signal allowed for precise measurements of $E_b = 251 \pm 4$ meV for BA2PbI4 and $E_b = 223 \pm 3$ meV for PEA2PbI4 (**see Section 6.1**). [124] This technique was then applied to 31 unique HP compositions in a subsequent article revealing an extremely strong correlation between $E_g$ and $E_b$. [5, 143]
- Miyata et al. resolved Landau levels associated transitions in the MAPbI3 magnetoabsorption spectrum, allowing for a measurement of $E_b = 16 \pm 2$ meV in the low-temperature orthorhombic phase (**see Section 7.2**). These experiments have since been repeated on various other 2D HP [103, 125] and 3D HP [100, 144] compositions.

**Table 3**: Exciton binding energy measurements on the most commonly studied lead-iodide perovskite compositions

| Material | Sample details | Technique | $E_b$ (meV) | Ref. |
|---|---|---|---|---|
| 3D HPs | | | | |
| MAPbI3 | Thin film (78 K) | Abs., fit | 17 | Niedzwiedzki et al. [6] |
| MAPbI3 | Thin film (RT) | Abs., fit | 13 | Yang et al. [136] |
| MAPbI3 | Thin film (RT) | Abs., fit | 9 | Yang et al. [135] |
| MAPbI3 | Thin film (170 K) | Abs., fit | 25 ± 3 | Saba et al. [145] |
| MAPbI3 | Thin film (14 K) | Abs., fit | 21 | Davies et al. [146] |
| MAPbI3 | Thin film (80 K) | Abs., fit | 34 ± 3 | Sestu et al. [147] |
| MAPbI3 | Thin film (4 K) | Abs., fit | 22 | Ruf et al. [148] |
| MAPbI3 | Thin film (13 K) | Abs., fit | 29 | Yamada et al. [149] |
| MAPbI3 | Thin film (RT) | EA | 7.4 ± 2 | Ziffer et al. [33] |
| MAPbI$_3$ | Thin film (50 K) | EA | 26 | Ruf et al. [113] |
| MAPbI$_3$ | Thin film (2 K) | Magnetoabsorption | 16 ± 2 | Miyata et al. [101] |
| MAPbI$_3$ | Single crystal (4.2 K) | Magnetoabsorption | 37 | Hirasawa et al. [116] |
| MAPbI$_3$ | Thin film (4.2 K) | Magnetoabsorption | 50 | Tanaka et al. [117] |
| MAPbI$_3$ | Single crystal (4 K) | Magnetoreflectance | 16 ± 2 | Yang et al. [150] |
| MAPbI$_3$ | Single crystal (1.5 K) | Magnetoreflectance | 3.1 ± 0.5 | Yamada et al. [151] |
| MAPbI$_3$ | Thin film | Temp-dependent PL | 19 ± 3 | Sun et al. [119] |
| MAPbI$_3$ | Thin film | Temp-dependent PL | 45 | Ishihara [152] |
| MAPbI$_3$ | Thin film | Temp-dependent PL | 32 ± 5 | Savenije et al. [120] |
| MAPbI$_3$ | Thin film | Temp-dependent PL | 32 ± 5 | Savenije et al. [120] |
| MAPbI$_3$ | Thin film | Temp-dependent PL | 7 – 115 | Niedzwiedzki et al. [6] |
| 2D HPs | | | | |
| BA$_2$PbI$_4$ | Single crystal (78K) | 2P-PLE | 180 | Chen et al. [126] |
| BA$_2$PbI$_4$ | Nanosheet (78K) | 2P-PLE | 190 | Chen et al. [126] |
| BA$_2$PbI$_4$ | Single crystal (5K) | 2P-PLE | 400 | Ziegler [153] |
| BA$_2$PbI$_4$ | Single crystal (RT) | Abs., fit | 482 | Chen et al. [154] |
| BA$_2$PbI$_4$ | Single crystal (1.5K) | Abs., fit | 320 ± 30 | Ishihara [121] |
| BA$_2$PbI$_4$ | Single crystal (78 K) | Abs., PLE | 260 | Zhang et al. [155] |
| BA$_2$PbI$_4$ | Nanosheet (78 K) | Abs., PLE | 260 | Zhang et al. [155] |
| BA$_2$PbI$_4$ | Thin film (50K) | EA | 220 | Amerling et al. [156] |
| BA$_2$PbI$_4$ | Thin film (<45 K) | EA | 251 ± 4 | Hansen et al. [124] |
| BA$_2$PbI$_4$ | Thin film (RT) | EA | 205 ± 7 | Hansen et al. [124] |
| BA$_2$PbI$_4$ | Nanosheet (4K) | PLE, Rydberg | 470 | Blancon et al. [104] |
| BA$_2$PbI$_4$ | Nanosheet (5K) | Refl., Rydberg | 490 | Yaffe et al. [14] |
| BA$_2$PbI$_4$ | Single crystal | Temp-dependent PL | 279 ± 46 | Chen et al. [157] |
| PEA$_2$PbI$_4$ | Single crystal (78K) | 2P-PLE | 160 | Chen et al. [126] |
| PEA$_2$PbI$_4$ | Single crystal (10K) | Abs., fit | 220 ± 10 | Hong et al. [19] |
| PEA$_2$PbI$_4$ | Single crystal (78 K) | Abs., PLE | 200 | Zhang et al. [155] |
| PEA$_2$PbI$_4$ | Nanosheet (78 K) | Abs., PLE | 200 | Zhang et al. [155] |
| PEA$_2$PbI$_4$ | Thin film (50 K) | EA | 190 ± 10 | Zhai et al. [158] |
| PEA$_2$PbI$_4$ | Thin film (<45K) | EA | 223 ± 3 | Hansen et al. [124] |
| PEA$_2$PbI$_4$ | Thin film (RT) | EA | 222 ± 10 | Hansen et al. [124] |
| PEA$_2$PbI$_4$ | Single crystal (2K) | Magnetoreflection | 265 | Dyksik [103] |
| PEA$_2$PbI$_4$ | Thin film | Temp-dependent PL | 453 | Chakraborty [159] |

## 3.4 Organic semiconductors

Organic semiconductors consist of pi-conjugated molecules, where conjugation (overlapping of p-orbitals) can lead to somewhat delocalized band structure within both small molecule and polymer systems. However, the degree of wavefunction overlap between adjacent molecules is typically small and is sensitive to the morphology of the molecular matrix, whether disordered or crystalline. As a result, bandwidths of molecular systems are tiny in comparison to inorganic systems (only 100 – 300 meV [160, 161]) and the fundamental band gap has essentially no optical signature, which makes measurements of $E_b$ quite challenging. Still, exciton effects are strong in organic semiconductors due to the weakly-screened Coulomb potential ($\varepsilon \sim 3$) and the value of $E_b$ is an important parameter to know for the engineering of organic optoelectronic devices.

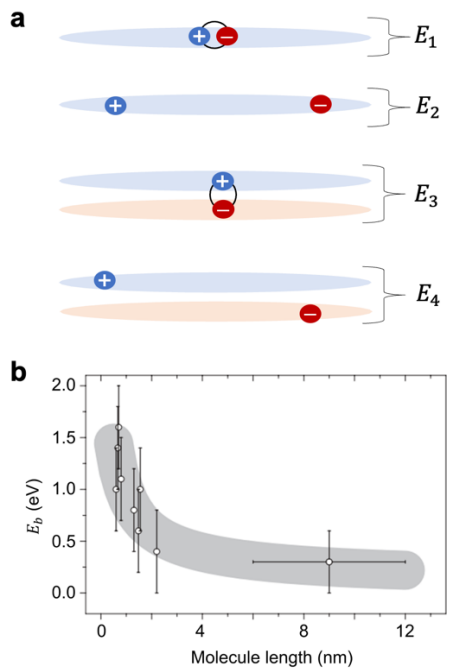

**Figure 5: (a)** Four different configurations for excited charges within a molecular system: (1) an exciton on a single molecule, (2) free electron and free hole on a single molecule, (3) an exciton at the junction of neighboring molecules (charge-transfer exciton) and (3) a free electron and free hole of neighboring molecules. For a Frenkel exciton $E_b = E_4 - E_1$, whereas $E_b = E_4 - E_3$ for a charge-transfer exciton. **(b)** Exciton binding energies of various organic semiconductors as a function of molecular length. Adapted from ref. [161] with permission from Springer, Copyright 2003.

Excitons in organic semiconductors come in three varieties: Frenkel, Wannier, and charge-transfer (CT), with the Frenkel type being the most common. [162] Strictly speaking, each type of exciton has a unique definition for its binding energy. In each case, it is important to give $E_b$ the definition that is most useful and relevant to device physics; that is, the difference between the energy required to create separated electron-hole pairs (known as the transport or electronic band gap), and the energy required to create bound electron-hole pairs (known as the optical band gap, i.e. the energy onset of optical absorption). In **Figure 5a**, four molecular configurations are depicted: (1) a bound electron-hole pair on a single molecule, (2) a free electron and a free hole on a single molecule, (3) an electron on one molecule bound to a hole on a neighboring molecule, and (4) a free electron on one molecule and a free hole on the neighboring molecule. For a Frenkel exciton on a small molecule or on a polymer lacking delocalized bands, configuration 2 cannot exist and the most useful definition is $E_b = E_4 - E_1$ [161]. Naturally, for a CT exciton the bound-state is given by configuration 3 rather than 1, and the exciton binding energy is $E_b = E_4 - E_3$ (see **ref.** [161], [163], [164], and [165] for useful discussions on these definitions). Wannier excitons only exist when the intermolecular wavefunction overlap is large, leading to conduction and valance bands with large bandwidths. In such cases, the picture of individual molecular units breaks down $E_b$ retains it standard definition (discussed previously in **Section 2**).

In most $E_b$ measurements on organic semiconductors in the literature, one assumes that the excitons are Frenkel or CT in nature and $E_4$ is measured by the transport gap. As previously stated, this band gap energy has essentially no optical signature in organic semiconductors and researchers are forced to resort to creative techniques to measure $E_4$ such as STS, [166] the combination of inverse photoelectron spectroscopy (IPES) and ultraviolet photoelectron spectroscopy (UPS), [162, 167-169] and even the external quantum efficiency (EQE) of a solar cell. [170] Unfortunately, these spectra typically have broad features with linewidths on the order of 100 meV and this poor spectral resolution translates to a high uncertainty in $E_b$. The energy of the bound-state configuration ($E_1$ or $E_3$), on the other hand, is the optical gap and can be easily obtained from the exciton peak in the absorption spectrum; the absorption spectra of organic semiconductors are typically dominated by excitons and their vibronics (see polydiacetylene absorption in **Figure 1d**).

Although it is rarely done in practice, the length of the molecule should be included when reporting $E_b$ measurements for organic semiconductors. In **Figure 5b**, the $E_b$ measurements of various organic semiconductors from the literature are plotted as a function of molecular length. The gray region represents a relationship derived by Knueffer, $E_b \approx e^2 \ln(2l/d)/\varepsilon_0 \varepsilon_r l + C$, where $l$ is the molecule length, $d$ is the molecule's diameter, and $C$ is a constant close to 0.1 eV (see ref. [161] for the full expression). This relationship between $E_b$ and $l$ can be understood by the fact that the HOMO and LUMO orbitals harboring the charge are delocalized over the length of a molecular unit, and thus, the Coulomb energy associated with the separated charges in configuration 4 of **Figure 5a** is a direct function of the molecule's size. A similar relationship between $E_b$ and $l$ was obtained by Hummer et al. using first-principle calculations on the first four linear oligoacenes (benzene through tetracene). [171] However, once the conjugation length of the molecule exceeds ~15 monomer units, $E_b$ reaches its long-chain-limit and it is no longer necessary to qualify $E_b$ as being specific to a given length. [172-174] Polaronic effects are another factor that must be considered when reporting $E_b$ of organic systems. In most organic semiconductors, conduction band electrons relax into polaronic states within ~100 fs after excitation. As a result, the transport gap is lower than the CBM-VBM separation (i.e., $E_g$) by as much as ~100 – 200 meV and $E_b$ is accordingly lower by the same amount. [163, 175] Thus, this effect should be considered when comparing $E_b$ as determined by the transport gap (as is the case for STS) vs. $E_b$ determined by the fundamental band gap (as is the case for electroabsorption).

*Unique challenges:*
- It is extremely rare for the interband transition or Rydberg states to have optical signatures.
- Absorption spectra are typically dominated by fairly broad peaks associated with Frenkel excitons and their vibronic satellites.

*Notable measurements:*
- Electroreflectance of polydiacetylene single crystals by G. Weiser (**see section 6.1**). [176, 177]
- Two-photon photoemission of pentacene single crystals by Muntwiler et al. The 1s, 2s, and 3s Rydberg states were resolved. [178]

**Table 4**: Exciton binding energy measurements on selected organic semiconductors

| Material | Notes | Technique | $E_b$ (meV) | Ref. |
| --- | --- | --- | --- | --- |
| α-6T | Thin film | UPS/IPES | 400 | Hill et al. [162] |
| α-6T | TF (CT exciton) | EA | 400 | Blinov et al. [179] |
| α-NPD | Thin film | UPS/IPES | 1000 ± 400 | Hill et al. [162] |
| Alq3 | Thin film | UPS/IPES | 1400 ± 400 | Hill et al. [162] |
| Alq3 | TF | UPS/IPES | 400 | Krause et al. [167] |
| CuPc | Thin film | UPS/IPES | 600 ± 400 | Hill et al. [162] |
| CuPc | TF | UPS/IPES | 100 | Krause et al. [167] |
| CuPc | TF | UPS/IPES | 440 ± 200 | Zahn et al. [169] |
| F8BT/PFB blend | TF | PL | ≥250 | Gélinas et al. [180] |
| HBC | TF | STS | 400 | Proehl [181] |
| P3DT | Disordered TF | Electroabsorption | 600 | Liess et al. [182] |
| P3HT | TF in solar cell | EQE | 730 ± 140 | Li et al. [170] |
| PCDTBT | TF in solar cell | EQE | 740 ± 190 | Li et al. [170] |
| PCE10 | TF in solar cell | EQE | 660 ± 150 | Li et al. [170] |
| PDA-4BCMU | Disordered TF | Electroabsorption | 600 | Liess et al. [182] |
| PDA-DCHD | Single crystal | Electroreflectance | 478 | Weiser [177] |
| PDA-DCHD | Single crystal | Electroreflectance | 480 | Sebastian et al. [176] |
| PDA-PFBS | Single crystal | Electroreflectance | 515 | Weiser [177] |
| PDA-PTS | Single crystal | Electroreflectance | 509 | Weiser [177] |
| PDES | Disordered TF | Electroabsorption | 500 | Liess et al. [182] |
| Pentacene | Crystal (CT exciton) | 2PPE | 430 | Muntwiler [178] |
| PFO | TF | Transient absorption | 600 ± 200 | Zhao et al. [183] |
| PFO | TF (spin-coat) | STS | 300 ± 100 | Alvarado et al. [166] |
| PPE | Disordered TF | Electroabsorption | 600 | Liess et al. [182] |
| PPV | TF (singlet exciton) | Photocurrent | 400 | Marks et al. [184] |
| PPV | TF | STS | 400 ± 100 | Kemerink et al. [172] |
| PPV-DOO | Disordered TF | Electroabsorption | 700 | Liess et al. [182] |
| PPV-DOO | TF | Transient absorption | 800 ± 200 | Zhao et al. [183] |
| PPV-MEH | Disordered TF | Electroabsorption | 600 | Liess et al. [182] |
| PPV-MEH | Metal/polymer/metal | EA and internal photoemission | 200 | Campbell et al. [175] |
| PPV-MEH | TF in solar cell | EQE | 1190 ± 150 | Li et al. [170] |
| PPV-AOPV | TF (spin-coat) | STS | 360 ± 100 | Alvarado et al. [166] |
| PTB7 | TF in solar cell | EQE | 710 ± 150 | Li et al. [170] |
| PTCDA | Thin film | UPS/IPES | 600 ± 400 | Hill et al. [162] |
| PTCDA | TF | UPS/IPES | 300 | Krause et al. [167] |
| PTCDA | TF | UPS/IPES | 440 ± 200 | Zahn et al. [169] |
| PTCDI | TF | UPS/IPES | 200 ± 200 | Zahn et al. [169] |
| PTV | Disordered TF | Electroabsorption | 600 | Liess et al. [182] |
| s-(CH)x | Disordered TF | Electroabsorption | 500 | Liess et al. [182] |

a)α-6T = α-sexithiophene; α-NPD = α-N,N'-diphenyl-N,N'-bis(1-naphthyl)-1,1'biphenyl-4,4'' diamine; Alq$_3$ = tris(8-hydroxy-quinoline)aluminum; CuPc = copper phthalocyanine; F8BT/PFB = polymer blend, structures specified in ref. [180]; HBC = Hexa-peri-hexabenzocoronene; P3DT = poly(3-alkylthiophene); P3HT = poly(3-hexylthiophene-2,5-diyl); PCDTBT = poly[N-9″-hepta-decanyl-2,7-carbazole-alt-5,5-(4′,7′-di-2-thienyl-2′,1′,3′-benzothiadiazole); PCE10 = poly[4,8-bis(5-(2-ethylhexyl)thiophen-2-yl) benzo[1,2-b;4,5-b']-dithiophene-2,6-diyl-alt-(4-(2-ethylhexyl)-3-fluorothieno[3,4-b]-thiophene-)-2-carboxylate-2-6-diyl)]; PDA-4BCMU = poly[5,7-dodecadiyn-1,12-diol-bis (n-butoxycarbonylmethyl urethane)]; PDA-DCHD = poly- 1,6-di(n-carbazolyl)-2,4-hexadiyne; PDA-PFBS = poly[2, 4-hexadiyne-1, 6-diol-bis(p-fluorobenzene sulfonate)]; PDA-PTS = poly[2, 4-hexadiyne-1, 6-diol-bis(p-toluene sulfonate)]; PDES = poly(diethynyl silane); PFO = poly(9,9'-dioctylfluorene); PPE = poly(p-phenylene ethylene); PPV = poly(p-phenylene-vinylene); PPV-DOO = poly(2,5-dioctyloxy-1,4-phenylenevinylene); PPV-MEH = Poly[2-methoxy-5-(2-ethylhexyloxy)-1,4-phenylenevinylene]; PPV-AOPV = poly[(2-methoxy-5-dodecyloxy)-1,4-phenylenevinyleneco-1,4-phenylenevinylene]; PTB7 = poly{4,8-bis[(2-ethylhexyl) oxy] benzo[1,2-b:4,5-b′]dithiophene-2,6-diyl-alt-3-fluoro-2-[(2-ethylhexyl)carbonyl]thieno[3,4-b]thiophene-4,6-diyl}; PTCDA = 3,4,9,10-perylenetetracarboxylic dianhydride; PTV = poly(thienylenevinylene); s-(CH)x see **ref.** [182] for structure.

## 4. Temperature-dependent PL

The goal of this technique is to measure quenching of the free-exciton PL intensity with increasing temperature, and then fit the decay in PL intensity to an Arrhenius model to obtain an ionization energy which is then identified as $E_b$. In the following, we derive the relationship between $E_b$ and the PL intensity as a function of temperature. Despite the widespread use of this technique in the literature, we find that it often produces inaccurate $E_b$ values. Therefore, assumptions inherent to the derivation that are potentially problematic are marked in bold in **Section 4.1** and these are discussed further in **Section 4.2**. Example Arrhenius fits for 2D HPs are shown, and finally, we present data showing that the $E_b$ measurement is sensitive to the fit procedure.

### 4.1 Derivation:

An excited population of excitons $n(t)$ will evolve according to:

$$\frac{\partial n(t)}{\partial t} = -\frac{n(t)}{\tau} \tag{7}$$

Solving for $n(t)$, one obtains:

$$n(t) = N_0 e^{-t/\tau} \tag{8}$$

The exciton population's decay rate, $1/\tau$ includes all decay pathways, both radiative ($R$) and nonradiative ($NR$).

$$\frac{1}{\tau} = \frac{1}{\tau_R} + \frac{1}{\tau_{NR}} \tag{9}$$

On the nonradiative side, the decay rate can include contributions from a wide range of decay mechanisms: thermal disassociation, trap states, thermal escape of carriers, [185] multiphonon relaxation, [186, 187], Auger nonradiative recombination, [188] and potentially other decay pathways as well.

$$\frac{1}{\tau_{NR}} = \frac{1}{\tau_{diss}} + \frac{1}{\tau_{trap}} + \frac{1}{\tau_{esc}} + \frac{1}{\tau_{phonon}} + \frac{1}{\tau_{Auger}} + \cdots \tag{10}$$

Due to the fact that each nonradiative decay pathway involves its own activation energy and rate, as is shown below, the number of fit parameters required to extract $E_b$ will ultimately be $2N$ where $N$ is the number of terms included in **Equation (10)**. Thus, in order to arrive at an equation that can be fit with a reasonable number of fit parameters, typically one assumes for a given range of temperatures that thermal disassociation is the predominant nonradiative decay pathway, i.e. $1/\tau_{NR} = 1/\tau_{diss}$ (**assumption 1**), and **Equation (10)** becomes:

$$\frac{1}{\tau} = \frac{1}{\tau_R} + \frac{1}{\tau_{diss}} \qquad (11)$$

Since only the radiative pathway is visible in the PL intensity, the exciton PL intensity as a function of time is given by (**assumption 2**):

$$I_{PL}(t) = \frac{n(t)}{\tau_R} \qquad (12)$$

The detector is sensitive to the intensity integrated over time.

$$\int_0^\infty I_{PL}(t)\, dt = \int_0^\infty \frac{n(t)}{\tau_R} dt = \int_0^\infty \frac{N_o e^{-t/\tau}}{\tau_R} dt = \frac{N_o}{\tau_R}\tau \qquad (13)$$

Plugging in **Equation (11)** results in:

$$I_{PL}(T) = \frac{N_o}{1 + \frac{\tau_R}{\tau_{diss}}} \qquad (14)$$

Since the exciton population at a given temperature is proportional to $e^{-E_b/k_B T}$ where $E_b$ is the exciton binding energy, the rate of dissociation $1/\tau_{diss}$ decreases with temperature according to $1/\tau_{diss} = 1/\tau_d\, exp(-E_b/k_B T)$ (**assumption 3**) where $1/\tau_d$ is the exciton dissociation *rate constant*. As such, $I_{PL}(T)$ can be rewritten as:

$$I_{PL}(T) = \frac{N_o}{1 + A_1 e^{-E_b/k_B T}} \qquad (15)$$

where $A_1 = \tau_R/\tau_d$. Practically speaking, **Equation (15)** rarely produces a satisfactory fit over the entire temperature range, and researchers can choose to either fit over a subset of the temperature range, or else include an additional non-radiative decay pathway (typically with a trap-state interpretation). Inclusion of another term in **Equation (11)** produces the following equation:

$$I_{PL}(T) = \frac{N_o}{1 + A_1 e^{-E_b/k_B T} + A_2 e^{-E_a/k_B T}} \qquad (16)$$

where $E_a$ is commonly interpreted as the trap-state activation energy. The $I_{PL}(T)$ dataset can be normalized such that $N_o = 1$, which leaves $A_1$, $E_b$, $A_2$, and $E_a$ to be adjusted as fit parameters in a least-squares fitting sequence.

### 4.2 Assumptions and challenges:

As indicated, there are three assumptions in the above derivation which may or may not be valid depending on the material system in question. Assumption 1 is that the exciton has only one significant nonradiative decay pathway (in the case of **Equation (15)**), or two significant nonradiative decay pathways (in the case of **Equation (16)**). In actuality, there may be more than two nonradiative recombination pathways that are significant. However, the inclusion of multiple pathways mandates additional fit parameters and this often produces untrustworthy fit results. While it is true that satisfactory fits over a finite range of temperatures can be obtained with **Equation (15)** (i.e. with only two fit parameters), selecting the bounds for the range is somewhat arbitrary and, in our experience, the fit results are often quite sensitive to this selection. In the case that multiple decay pathways are included, such as in **Equation (16)**, it is not uncommon for the least-squares fit to produce activation energies of similar magnitude, rendering their physical origins unclear. For example, if the values of $E_b = 100$ meV and $E_a = 50$ meV are obtained, it is difficult to know which is the supposed exciton binding energy and which is the trap-state energy.

Assumption 2 is that $n(t)$ in **Equation (12)** represents a real population of excitons. In reality, it has been shown that free carriers can also contribute to "exciton" photoluminescence due to many-body interactions, especially in low-$E_b$ systems at high carrier densities.[189-191] Although this is a topic of some controversy,[189, 192] experimental results indicate that free carrier contributions to the 1$s$ exciton PL peak are negligible when the material is excited with reasonably low photon densities (<$10^{10}$ photons/cm$^2$),[192] and therefore, the validity of assumption 2 can be ensured by exciting with a continuous wave laser or a pulsed laser at low-to-moderate fluences. Related to this problem, however, is a more practical challenge of accurately separating the 1$s$ exciton peak from neighboring spectral features. For some materials, neighboring peaks from various bound exciton states overlap with the free 1$s$ exciton emission and non-trivial fitting is needed to extract the integrated intensity of the free exciton emission. In many cases, accurate separation of the peaks is not possible over a wide temperature range.

Finally, assumption 3 is that the exciton's radiative lifetime ($\tau_R$) is temperature independent. In reality, it is quite common for this assumption to be violated, as time-resolved photoluminescence (TRPL) often measures a change in $\tau_R$ with temperature.[115, 193-195] In some instances, the temperature dependence of $\tau_R$ can be determined with TRPL and the functional form of $\tau_R(T)$ can be incorporated into **Equation (15)** and **Equation (16)**.[196] For example, if $\tau_R$ increases approximately linearly with temperature ($\tau_R =$ $\tau_{R0}T$), then $A_1$ can be replaced with $a_1T$ where $a_1 = \tau_{R0}/\tau_{diss}$. In some cases, however, the functional form of $\tau_R(T)$ is far more complex [194, 195] and in these instances one should avoid interpreting Arrhenius activation energies as $E_b$.

### 4.3 Application to HPs

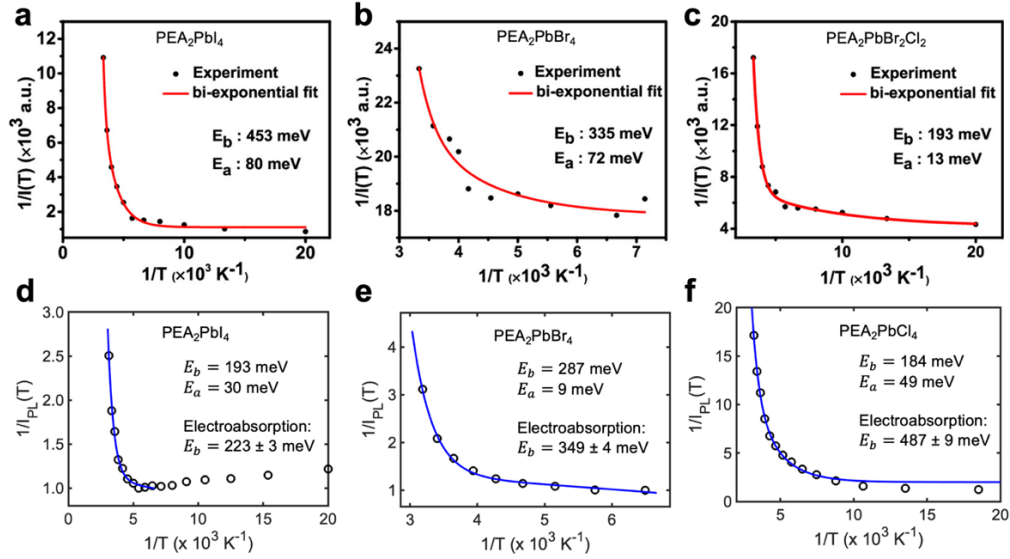

**Figure 6: (a)-(c)** Arrhenius model fitting of the temperature dependent PL, plotted as the inverse of the integrated PL intensity vs 1/T, for (a) PEA$_2$PbI$_4$ (b) PEA$_2$PbBr$_4$ and (c) PEA$_2$PbCl$_4$. Reprinted from ref. [159] with permission from the American Chemical Society, Copyright 2020. **(d) – (f)** Original data from our group attempting to reproduce the results in panels (a) – (c). Note that the 2D HP composition in panel (f) is slightly different from that in panel (c), namely PEA$_2$PbCl$_4$ vs PEA$_2$PbBr$_2$Cl$_2$. Included in the Figure is the known value of $E_b$, obtained from Franz-Keldysh and Stark features in the electroabsorption spectrum. [5] The value obtained for PEA$_2$PbBr$_2$Cl$_2$ via electroabsorption is $E_b = 449 \pm 8$ meV.

Both **Equation (15)** and **Equation (16)** have been applied to fit the temperature-quenching of PL intensity for a wide range of material systems, e.g. III-V semiconductors, [196-198] CdTe and CdSe quantum dots, [199-201] 3D HPs, [118-120, 195, 202-204] 2D HPs, [115, 121, 122, 159, 205, 206] and more. While some studies are aware of the above assumptions and are appropriately cautious in assigning the fitted Arrhenius activation energy to $E_b$, other studies are not and subsequently use the fitted $E_b$ value to make further claims within the context of the study. In particular, the use of these equations has been widespread in studies on HPs, a system where knowledge of $E_b$ is critical for many optoelectronic applications but measurements of $E_b$ are challenging.

In **Figure 6a-c**, we highlight one such study wherein $I_{PL}(T)$ was fit to **Equation (16)** for three 2D HP compositions, in order to determine the effect of the halide atom (I,

Br, or Cl) on $E_b$.[159] The fits accurately match the experimental data for a selected range of temperatures, and $E_b$ values are reported for PEA$_2$PbI$_4$ (453 meV), PEA$_2$PbBr$_4$ (335 meV), and PEA$_2$PbBr$_2$Cl$_2$ (193 meV). In an effort to reproduce these results, our group independently measured $I_{PL}(T)$ for PEA$_2$PbI$_4$, PEA$_2$PbBr$_4$, and PEA$_2$PbCl$_4$; these results are shown in **Figure 6d-f**. We obtained similar $E_b$ results in the case of PEA$_2$PbBr$_4$ (287 meV) and PEA$_2$PbCl$_4$ (184 meV), however, a much lower value of 193 meV was measured for PEA$_2$PbI$_4$. This disparity is likely due to a difference in the density and nature of trap states, which have been shown to be sensitive to the specifics of the PEA$_2$PbI$_4$ processing conditions.[207] Nevertheless, none of Arrhenius-determined $E_b$ values correlate with the known binding energy obtained from Stark and Franz-Keldysh features in the electroabsorption spectrum (see **Section 6.1**). In fact, the Arrhenius-determined $E_b$ values are opposite the trend determined by electroabsorption, which is that $E_b$ unambiguously decreases moving from PEA$_2$PbCl$_4$ (487 ± 9 meV) to PEA$_2$PbBr$_4$ (349 ± 4 meV) to PEA$_2$PbI$_4$ (223 ± 3 meV).

Without further investigation into each of the three assumptions discussed above, it is difficult to know why the biexponential fits in **Figure 6** fail to produce the correct exciton binding energy, and herein lies the central challenge with using Arrhenius fits to determine $E_b$. That is, investigating the degree to which the material satisfies assumptions 1-3 is very non-trivial. In fact, it is probably easier in every case to measure $E_b$ using an alternative, more trustworthy method than to verify assumptions 1-3.

We do however note that, in general, the Arrhenius-determined $E_b$ values reported throughout the literature tend to be correct to within an order of magnitude. For example, for 3D HPs, Arrhenius activation energies fall in the range of ~20 to 60 meV [118-120, 195, 202-204] which is higher than the known $E_b$ values (7 – 30 meV) [33, 100, 101], but accurate in the sense that they are consistently smaller than Arrhenius activation energies of 2D HPs (~ 50 – 500 meV).[115, 121, 122, 159, 205] Along this vein, Arrhenius fits tend to get the order-of-magnitude of $E_b$ correct for III-V systems [196-198] and other systems as well.

### 4.4 Sensitivity of fit procedure

When performing a fit according to **Equation (15)** or **Equation (16)**, one is also faced with the decision of which $\{x,y\}$ variables to select as inputs to the least-squares regression. Surveying the literature, all permutations of $I_{PL}$ and $I_{PL}^{-1}$ as functions of $T$ and $1/T$ are commonly used. While the selection of $T$ vs $1/T$ has no effect on the final value of the fit parameters (least-squares procedures are typically based on the vertical offsets), the same is not the case for $I_{PL}$ vs $I_{PL}^{-1}$. In **Figure 7**, the same PEA$_2$PbBr$_4$ dataset featured previously in **Figure 6e** is plotted with three different y-axis variables: $I_{PL}$ in **Figure 7a**, $I_{PL}^{-1}$ in **Figure 7b**, and $\ln(I_{PL})$ in **Figure 7c**. Here, the single exponential form of $I_{PL}(T)$ from **Equation (15)** is selected as the fitting function for simplicity. In each case, the fit was performed with the specified y-axis variable ($I_{PL}$, $I_{PL}^{-1}$, or $ln(I_{PL})$) as inputs to the least-squares procedure; and as is clear from the fitted values reported in **Figure 7**, the fit results are quite sensitive to this choice.

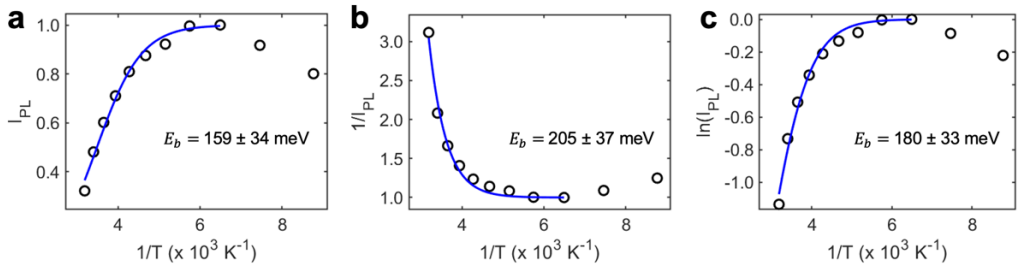

**Figure 7**: The PEA$_2$PbBr$_4$ dataset from **Figure 6e** plotted and fitted three ways: **(a)** The integrated PL intensity ($I_{PL}$) vs inverse temperature, (b) the inverse of the integrated PL intensity ($I_{PL}^{-1}$) vs inverse temperature, and finally, **(c)** the logarithm of the integrated PL intensity ($ln(I_{PL})$) vs inverse temperature. The black circles represent experimental data, the blue line is a least-squares fit according to **Equation (15)**. Errors in $E_b$ are reported as 95% confidence intervals for the least-squares fit.

Due to the exponential-nature of the relationship between the $I_{PL}$ and $T$, selecting $I_{PL}$ or $I_{PL}^{-1}$ gives greater weight to the large y values (low temperature for $I_{PL}$ and high temperature for $I_{PL}^{-1}$) within the least-squares procedure. As a result, selecting $I_{PL}$ underestimates $E_b$ while selecting $I_{PL}^{-1}$ overestimates $E_b$ for this dataset. Taking the natural logarithm of $I_{PL}$ leads to a more even weighting all data points. Thus, despite $ln(I_{PL})$ being the least-common selection for $y$ variables in the literature, it is the optimal choice for this situation.

## 5. Resolving the Rydberg Series

### 5.1 Probing bright states with one-photon transitions

Optically detecting higher members of the exciton's Rydberg series is conceptually quite straightforward in comparison to many of the other techniques discussed in this article, although the practical implementation can be challenging. As described in **section**

**2**, an exciton's absorption spectrum includes peaks at the eigenenergies corresponding to the $s$ state excitons. In accordance with the Bohr model and **Equation (3)**, the absorption strength of each peak is attenuated by a factor of $n^{-3}$ relative to the 1$s$ state and the peak energies become closely spaced as they approach $E_g$:[8, 17]

$$E_{n,3D} = E_g - \frac{R}{n^2} \qquad (17)$$

where $R$ denotes the Rydberg energy which is related to the binding energy of the 1$s$ state. In 3D, $E_b = R$ whereas in an ideal 2D system, $E_b = 4R$. In fact, each of the $E_{n,2D}$ energies are increased in two-dimensional systems, as follows: [17]

$$E_{n,2D} = E_g - \frac{R}{(n-1/2)^2} \qquad (18)$$

Thus, if a series of Rydberg peaks is resolved with energies $E_n$, this series can be fit according to Equation (17) or Equation (18). Extrapolating the linear fit to $n = \infty$ yields $E_g$, and the exciton binding energy then deduced by $E_b = E_g - E_{1s}$. If a series of excitonic peaks are resolved but they do not follow a pure Rydberg pattern as per Equation (17) or Equation (18), then the highest visible peak only allows an upper bound for $E_b$ to be established.

The absorption *strength* of excitons also increases moving from 3D → 2D as the oscillator strength of the $n$th state transitions from $\propto n^{-3}$ in three dimensions to $\propto (n-1/2)^{-3}$ in two dimensions. [17] Despite the overall enhancement of excitonic absorption in the 2D limit, the oscillator strengths and binding energies of the $n > 1$ states are weaker *relative to the* 1$s$ state. For example, in the 3D limit, the 2$s$ exciton's binding energy and oscillator strength are 0.25× and 0.125× that of the 1$s$ state, respectively; whereas in the 2D limit these factors of attenuation are 0.111 and 0.037, respectively.

For some materials, it is possible to obtain clear optical signatures of the Rydberg series by cooling extremely high-purity samples to liquid helium temperatures for optical measurements such as transmission, reflection, or PLE spectroscopy. For example, in their 2014 study, Kazimierczuk et al. resolved the exciton's Rydberg series in copper oxide ($Cu_2O$) up to an astonishingly large Rydberg state of $n = 25$. [208] The researchers polished a natural $Cu_2O$ crystal down to a thickness of 34 μm and measured the transmission at 1.2 K using a tunable laser with spectral resolution of 5 neV. As shown in the high-resolution absorption spectrum in **Figure 8a**, the exciton peaks become closely spaced with diminishing oscillator strengths as they converge towards $E_g$. Incidentally, because the highest VB and lowest CB originate from the 3d and 4s Cu orbitals, respectively, which have the same parity, electric dipole transitions to create $s$-type excitons are forbidden and the excitons which are created are $p$-type. However, since the energy of a state is determined by the principle quantum number $n$, the $p$-type energy levels are the same as what has been discussed above for the $s$-type states, albeit the $n = 1$ exciton does not appear because there is no $1p$ state.

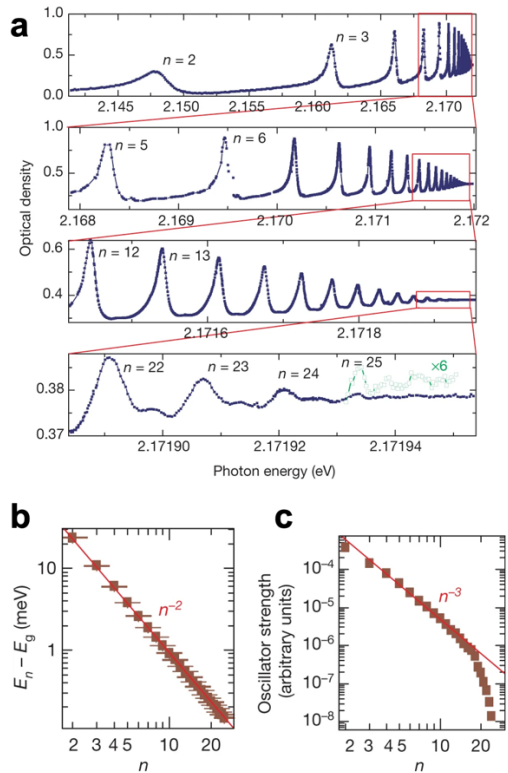

**Figure 8**: **(a)** A high-resolution absorption spectrum of the Cu$_2$O band edge at 1.2K. The peaks correspond to resonances of the excitons $np$-states. **(b)** The binding energy of each state. The trend exhibits a clear $n^{-2}$ dependence. **(c)** The oscillator strengths of each exciton peak. For $n > 18$, the measurments deivate from the expected $n^{-3}$ dependence due to exciton-exciton interactions. Reprinted from ref. [208] with permission from Springer Nature, Copyright 2014.

After a 400-fold magnification of the horizontal axis, the peak corresponding to the $n = 25$ state is visible on the bottom panel of **Figure 8a**. Each peak's energy, $E_g - E_n$, is plotted in **Figure 8b** and the trend follows the prediction of $n^{-2}$, as expected from **Equation(17)**, to a remarkable degree. The authors integrated the peak intensities to compare their relative oscillator strengths and the results closely match the theoretical $n^{-3}$ dependence up to $n = 18$ at which point there is a pronounced deviation from the $n^{-3}$ trendline (see **Figure 8c**) due to exciton-exciton interactions.[208] The authors extrapolated the energies to estimate $E_{1s}$ and $E_g$ and obtained $E_b = 92$ meV. Remarkably, this is the only $E_b$ measurement featured in this article where neither the $E_{1s}$ nor $E_g$ energy levels were directly determined.

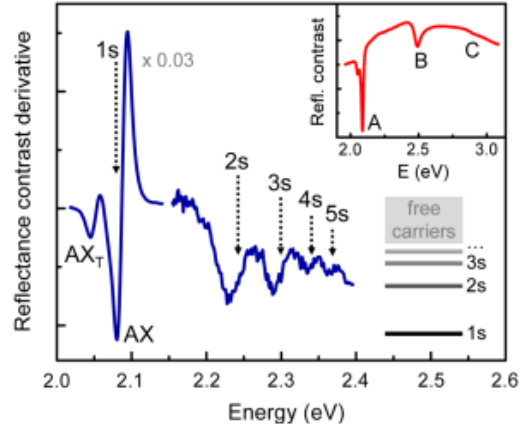

**Figure 9**: The derivative of the reflectance contrast, $(d/dE)\Delta R/R$, of a WS$_2$ monolayer on an Si/SiO$_2$ substrate. Reprinted from ref. [66] with permission from the American Physical Society, Copyright 2014.

Resolving higher members of the Rydberg series has also been a popular route towards measuring $E_b$ in TMDC monolayers and 2D HPs. In fact, the first $E_b$ measurements for TMDC monolayers were obtained in this manner in 2014, four years after the realization of monolayer MoS$_2$.[48, 54] Two studies probed the $p$-states via two-photon absorption (**Section 5.2**) and one, by Chernikov et al.,[66] resolved the $s$-series by measuring the reflection from a WS$_2$ monolayer. The WS$_2$ composition was selected due to the large ~0.4 eV separation of the A and B excitons, allowing the opportunity to resolve the A-exciton's Rydberg series between the broad A and B peaks. The authors exfoliated WS$_2$ monolayer was placed on an Si substrate coated with 300 nm of SiO$_2$ and the reflectance contrast, $\Delta R/R = (R_{sample} - R_{substrate})/R_{substrate}$ was measured, which is shown as the red line in the inset of **Figure 9**. The broad background is from interference between the Si and SiO$_2$ substrate layers, while the minima labeled A, B, and C correspond to well-known $1s$ exciton states corresponding to the spin-orbit split, nondegenerate band gaps at the K point.

The derivative spectrum, $d/dE\ \Delta R/R$ in **Figure 9**, exhibits multiple oscillations which the authors assign to the $1s$, $2s$, ..., $5s$ transitions. Interestingly, these energies do not follow the energy ladder predicted by the 2D Wannier model in **Equation (18)**, nor the 3D model of **Equation (17)**. From these equations, one would expect $\Delta_{21} = 0.75 E_b$ and $\Delta_{21} = 0.89 E_b$ and models, respectively. However, in **Figure 9**, the 2s energy is approximately the average of the 1s and 5s points, resulting in a situation much closer to $\Delta_{21} \approx 0.5 E_b$.

This non-hydrogenic ladder is attributed to the anisotropic screening of the Coulomb potential, where the dielectric medium of the bottom layer (SiO$_2$) is over twice that of the top layer (vacuum). The measured non-hydrogenic ladder is actually predicted by the Rytova-Keldysh potential of **Equation (6)**; the authors fit their data points according to this potential to estimate $E_g = 2.41$ eV and $E_b = 320 \pm 40$ meV. Since magneto-optical studies subsequently corroborated that these oscillations in the reflectance correspond to Rydberg states [92], this appears to be an accurate and important measurement of $E_b$ for TMDC monolayers, establishing the non-hydrogenic scaling of the exciton's Rydberg series in the presence of low dielectric media. However, given the low signal-to-noise in **Figure 9**, it would be helpful to see the measurement replicated for multiple samples and the statistical deviations analyzed. In the following, we showcase two examples of Rydberg series measurements for 2D HPs, also with low signal-to-noise, where the $E_b$ measurement did not bear out to be in as good of agreement with subsequent studies.

Two independent studies by Yaffe et al. in 2015 [14] and Blancon et al. in 2018 [104] reported optical detection of a Rydberg series for 2D HP single crystals with a composition of BA$_2$PbI$_4$. In the case of Yaffe et al., the measured optical signal is the absorption $A = 1 - T - R$ and reflectance contrast $\Delta R/R$. The derivatives of these spectra are displayed in **Figure 10a** for thin BA$_2$PbI$_4$ nanosheets and a bulk crystal. The top panel shows $dA/dE$ for nanosheet #1, the middle two panels show $(d/dE)\, \Delta R/R$ for nanosheets #2 and #3, and the bottom panel shows $R$ for the bulk crystal. After cooling to 4K, there is a coexistence of the low-temperature crystal phase (I) and high-temperature crystal phase (II) within each of the nanosheets. The 1s (I) and 1s (II) excitons, corresponding to the low- and high-temperature crystalline phases respectively, have clear resonances for each of the nanosheets. Above these resonances, there are several small oscillations (labeled $2s$ - $4s$ in **Figure 10a**), which the authors assign to $ns$ Rydberg series. While the respective 1s points are clear and energetically aligned to within ~ 5 meV, the $s$-state features are faint and vary significantly by ~ 50 meV between the different samples. The authors average the $s$-state energies and invoke a 2D hydrogen model to obtain $E_b = 490 \pm 30$ meV. A similar $E_b$ value was obtained for the same 2D HP perovskite composition by Blancon et al. [104] In this study, both the absorption and the PLE spectra were measured for a BA$_2$PbI$_4$ single crystal flake at 4 K; these spectra are depicted by the black (absorption) and red (PLE) lines in **Figure 10b**. Similar to the study by Yaffe et al., faint features in the spectra are assigned to the *s*-state Rydberg energies. The energies were selected from the minima in the second derivative of the absorption (**Figure 10c**), where the data points have been smoothed. Using a 2D hydrogen model, the values of $E_g = 3.016 \pm 0.011$ eV and $E_b = 472 \pm 82$ meV were obtained.

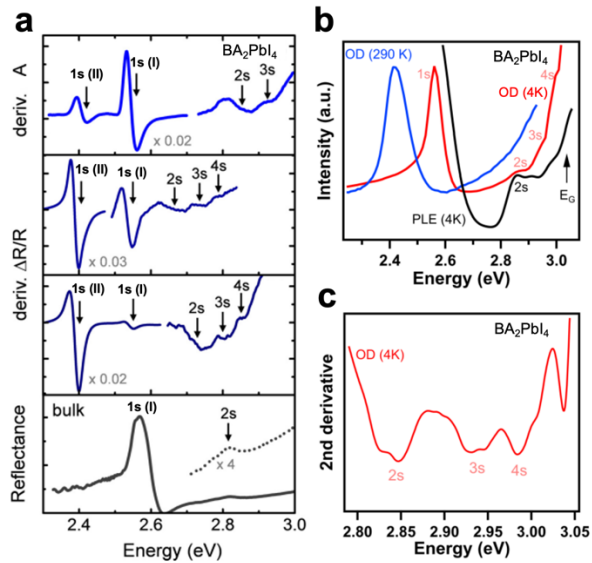

**Figure 10**: **(a)** Assignments of the exciton's Rydberg series for BA$_2$PbI$_4$. The top panel shows the derivative of a nanosheet's absorption spectrum. The middle two panels show the derivative of the "reflectance contrast" spectrum for two additional nanosheets. The bottom panel shows the reflectance of a bulk single crystal. Adapted from ref. [14] with permission from the American Physical Society, Copyright 2015. **(b)** The absorption spectrum of a BA$_2$PbI$_4$ single crystal flake measured at 290 K (blue), 4K (red) and again at 4K using PLE spectroscopy (black) **(c)** The second derivative of the red line in panel (b). Panels (b-c) are reprinted from ref. [104] with permission from Springer Nature, Copyright 2018.

While the studies by Yaffe et al. and Blancon et al. were important in advancing the understanding of excitons in 2D HPs, and include several contributions beyond their reporting of $E_b$ for BA$_2$PbI$_4$, these $E_b$ measurements have since been contradicted by electroabsorption, magnetoabsorption, and PLE measurements that show $E_b$ for lead-iodide 2D HPs is ~250 meV [103, 115, 124, 125, 155] and $E_g$ ~ 2.8 eV for BA$_2$PbI$_4$ in the low-temperature phase [124, 126, 155]. Incidentally, the interband transition at 2.8 eV is resolved in the 4K PLE spectrum in **Figure 10b**, where a step-like feature appears near this energy. Thus, when resolving Rydberg states, it is important the spectral features are considered en masse. If Rydberg states are assigned, the features should have a reasonable level of clarity, or in the case of low signal-to-noise, the measurements should be replicated for multiple samples. Both the samples and the optical measurements in the studies by Yaffe et al. and Blancon et al. were of the highest of quality, which begs the question of why these measurements did not bare out to be in agreement with subsequent electroabsorption, magnetoabsorption, and PLE experiments. It may be that for $BA_2PbI_4$, the above-gap exciton peaks from multivalley absorption are too broad and the overlap of these peaks with Rydberg features prevents their optical detection.[104, 138] This opens the question of whether it is possible to resolve $n > 1$ Rydberg features in any 2D HP sample, or if these features are firmly buried beneath the background of neighboring features regardless of sample quality and measurement technique.

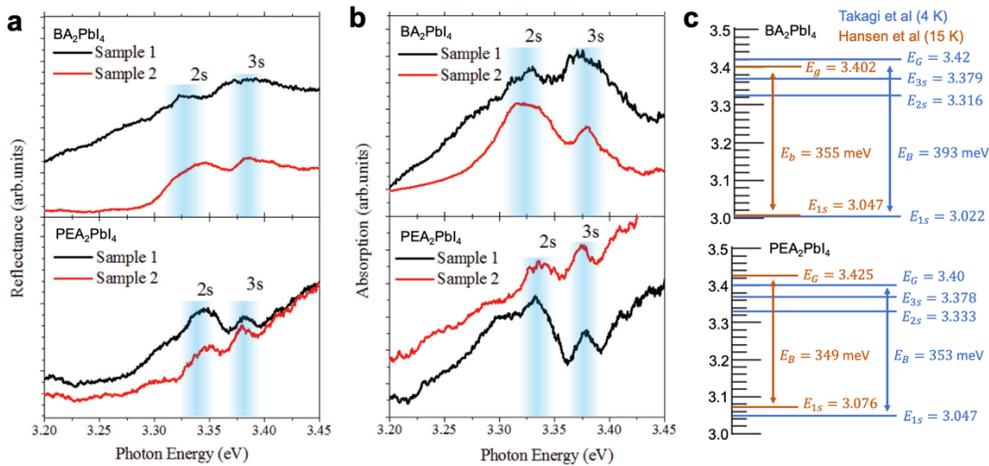

**Figure 11**: **(a)** The reflectance spectra of $BA_2PbI_4$ and $PEA_2PbI_4$ single crystals. **(b)** The absorption spectra of the same materials, obtained by a Kramers–Kronig transformation of the data in panel (a). Panels (a-b) are reprinted from ref. [209] with permission from the American Physical Society, Copyright 2013. **(c)** The Rydberg energy levels reported by Takagi et al. (blue) in comparison to the $E_{1s}$ and $E_g$ values obtained via electroabsorption measurements on the same compositions (orange).

Incidentally, in a 2013 study by Takagi et al., faint features between $E_{1s}$ and $E_g$ were resolved, assigned to the $s$-state Rydberg states, and reproduced for different samples.[209] These features are shown in **Figure 11a**, where the reflectance spectra for two 2D HP compounds, butylammonium lead bromide ($BA_2PbBr_4$) and phenethylammonium lead bromide ($PEA_2PbBr_4$) exhibit small peaks near 3.35 eV. The authors assign these peaks to the $2s$ and $3s$ states. For both $BA_2PbBr_4$ and $PEA_2PbBr_4$, the experiment is reproduced for two samples and the peak positions are in good agreement. Performing Kramers-Kronig analysis on the reflection data produces the absorption spectra shown in **Figure 11b**, where the $2s$ and $3s$ peaks are still faint, but nonetheless appear to be present at the same energies. Since only the $1s$ - $3s$ energies are measured, extrapolation to find $E_g$ is quite sensitive to selection the $2s$ and $3s$ energies as well as the model, i.e. 2D hydrogen, 3D hydrogen, or other. The authors choose to extrapolate the $2s$ and $3s$ points using a 2D hydrogen model and arrived at $E_b = 393$ meV for $BA_2PbBr_4$ and $E_b = 356$ meV for $PEA_2PbBr_4$. Unlike the Rydberg series measurements on $BA_2PbI_4$, these results are in excellent agreement with electroabsorption measurements where definitive values of $360 \pm 11$ meV and $349 \pm 4$ meV have been obtained for $BA_2PbBr_4$ and $PEA_2PbBr_4$, respectively. The Rydberg energy ladders reported by Takagi et al. are shown in **Figure 11c** side-by-side with the electroabsorption-determined values for $E_{1s}$ and $E_g$. While at first glance there appear to be some discrepancies, Takagi's measurements include moderate uncertainty from both the measured peak positions and the choice of 2D vs 3D Wannier model to extrapolate the trend to $E_g$. We calculated the $1\sigma$ uncertainty to be 24 meV, which may explain the 28 meV discrepancy between the $E_b$ values for $BA_2PbI_4$. Meanwhile, the $E_b$ values agree closely ($\pm$ 4 meV) for $PEA_2PbBr_4$.

## 5.2 Probing dark states with two-photon transitions

Most direct-gap semiconductors have dipole-allowed interband transitions. For these systems, one-photon transitions can only excite exciton states with even parity, such as the $s$-state Rydberg series, while two-photon transitions can reach exciton states with odd parity, such as the $p$-state Rydberg series. [63] Probing the $p$ states can be an experimental preference when the $s$-state series is difficult to resolve for $n > 1$. In terms of converting the $n > 1$ energies to $E_b$, it makes little difference whether the $s$ or $p$ states are measured because the energy of each states is determined solely by the principle quantum number, $n$.

Typically, two-photon PLE (2P-PLE) is used to resolve the $p$ series. In these experiments, the PL intensity is monitored while the sample is irradiated by an IR pulsed laser with tunable energy such that $2E_{photon}$ is between $E_{1s}$ and $E_g$. Similar to PLE, increased absorption will manifest as an increase in PL intensity, and the $p$-state series for $n \geq 2$ manifests as a series of peaks in the absorption according to **Equation (17)** and **Equation (18)**. The earliest $E_b$ measurements on TMDC monolayers came by way of 2P-PLE experiments. In two individual studies, the $1s$, $2p$, and $3p$ exciton peaks were reported for monolayer WSe$_2$ by He et al. [64] and for monolayer WS$_2$ by Ye et al. [63]

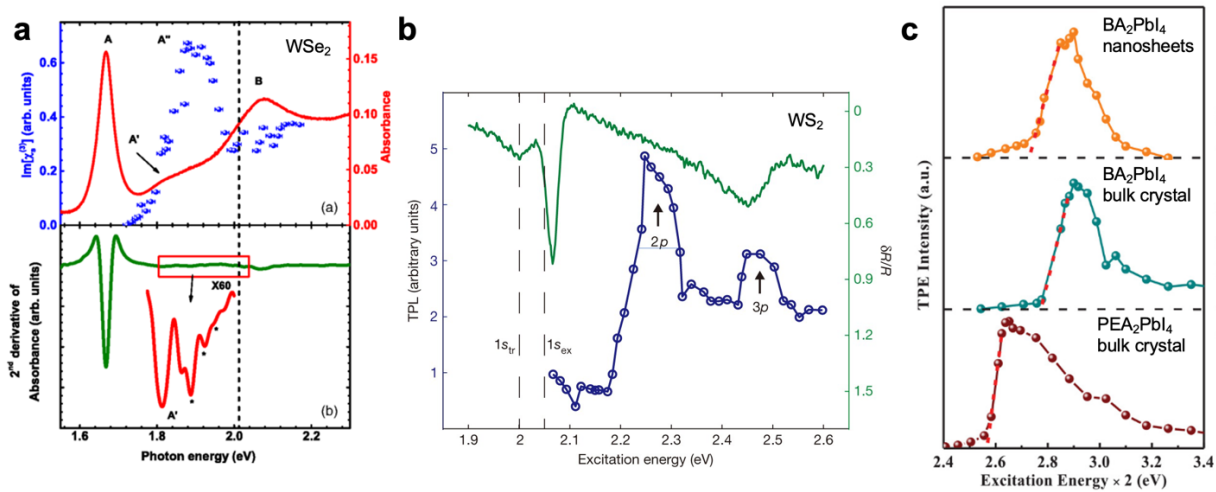

**Figure 12: Two photon absorption of TMDC monolayers and 2D HPs. (a)** The one-photon (red) and two-photon (blue) absorption spectra as measured by reflectance contrast and 2P-PLE spectroscopies, respectively, for an exfoliated WSe$_2$ monolayer on a quartz substrate. Reprinted from ref. [64] with permission from the American Physical Society, Copyright 2014. **(b)** The one-photon (green) and two-photon (blue) absorption spectra, also measured by reflectance contrast and 2P-PLE spectroscopies, of an exfoliated WS$_2$ layer on a quartz substrate. Reprinted from ref. [63] with permission from Springer Nature, Copyright 2014. **(c)** The two-photon absorption spectra of various 2D HP crystals: BA$_2$PbI$_4$ nanosheets (top), BA$_2$PbI$_4$ bulk crystals (middle) and PEA$_2$PbI$_4$ bulk crystals (bottom). Reprinted from ref. [126] with permission from Wiley-VCH, Copyright 2020.

As shown in the top panel of **Figure 12a**, the linear absorption of WSe$_2$ (red line) exhibits $1s$ exciton peaks for the typical A and B transitions. Between these peaks, the 2P-PLE signal (blue circles) shows a broad peak (labeled A'') where the $p$-states absorb. He et al. assign this peak to an overlap of the $2p$ and $3p$ excitons, but due to its broad nature, they choose not to assign $2p$ and $3p$ energies. Fortunately, higher members of the $s$-state series are resolved in the absorption. Similar to the study by Chernikov et al. on WS$_2$ (**Figure 9**), He et al. exfoliate the crystal on an Si/SiO$_2$ substrate down to a single monolayer, measure the absorption through the reflectance contrast $\Delta R/R$, and look for zero-crossings in the first derivative and dips in the second derivative that may correspond to the exciton's $s$-state energies. The second derivative spectrum is shown in the bottom panel of **Figure 12a**, where A' is assigned to the $2s$ state and the black dots mark the positions assigned to the $3s - 5s$ states. Although the features are faint, the authors are able to replicate the experiment for five samples and find the features to be reproducible in all five samples. Thus, in this measurement of $E_b$, the dark exciton states only play a minor role by increasing the confidence of the $1s - 5s$ assignments. The WSe$_2$ $s$-state series follows a similar non-hydrogenic energy ladder to that observed by Chernikov et al. for WS$_2$, and the authors obtain a binding energy of $E_b = 370$ meV, which is in good agreement (± 100 meV) with subsequent STS measurements. [85, 88]

An extremely similar dark-exciton spectrum was obtained in the same year for a WS$_2$ monolayer on a quartz substrate, by Ye et al., and is shown in **Figure 12b**. Here, the derivative of the reflectance contrast $d/dE$ $\Delta R/R$ (green) provides a zero-crossing at the $1s$ level while the 2P-PLE (dark blue) exhibits two peaks that are assigned to the $2p$ and $3p$ resonance energies. Unlike the studies by Chernikov et al. and He et al., the signal-to-noise ratio (SNR) of the reflectance contrast is insufficient to resolve the $s$-state Rydberg series and the authors must rely on the 2P-PLE spectrum to make assignments of the $n > 1$ states. The $2p$ peak in **Figure 12b** is slightly narrower than that resolved for WSe$_2$ by He et al., but the difference is slight (~80 meV vs ~100 meV). Furthermore, there is an additional peak in the WS$_2$ 2P-PLE spectrum near 2.5 eV which the authors assign to the $3p$ state and this motivates a high $E_b$ measurement of 700 meV.

Thus, we see in two separate studies that similar features in the 2P-PLE spectrum of TMDC monolayers were given different assignments— the $2p$ state in the case of Ye et al. and a grouping of the $2p$ and $3p$ states in the case of He et al, leading to conflicting $E_b$ measurements (700 meV vs. 370 meV, respectively). This discrepancy in measurement is too large to attribute to the material composition since WSe$_2$ and WS$_2$ have similar dielectric constants [210, 211] and also cannot be attributed to differences in dielectric environments given that both materials were placed between SiO$_2$ and air. These examples highlight the difficulty of making accurate assignments of the $n > 1$ exciton states with limited data and based off spectra with low signal-to-noise. In the case of He et al., the authors had the advantage of additionally resolving the $s$-state series (and reproducing this measurement in five samples) which led to an interpretation of the 2P-PLE peak that has borne out to be in better agreement with subsequent studies, seeing as STS and magneto-optical experiments indicate $E_b$ is ~300 - 450 meV for TMDC monolayers on $SiO_2$. [77, 85, 87-89, 92]

Finally, we turn our attention to the 2P-PLE spectra of 2D HPs measured by Chen et al.,[126] which is displayed in **Figure 12c**. In this study, single crystals of PEA$_2$PbI$_4$ and BA$_2$PbI$_4$ were exfoliated to varying thickness, and the 2P-PLE spectra of nanosheets ($t <$ 5 nm) and bulk crystals ($t >$ 5 nm) were measured. Unlike the 2P-PLE spectra of the TMDC monolayers, the authors assign the broad feature to the band edge and use a linear fit to extract $E_g$. There is an additional faint peak below the band onset (not shown here) that the authors assign to the $2p$ state.[126] The assignment of the band gap leads to an $E_b$ measurement of 160 meV for PEA$_2$PbI$_4$ bulk crystals, 180 meV for BA$_2$PbI$_4$ nanosheets, and 190 meV for BA$_2$PbI$_4$ bulk crystals. Interestingly, the measurements of $E_g$ in this study seem to be quite accurate seeing as they agree with electroabsorption and PLE measurements.[124, 155] However, the final reporting of $E_b$ is about 50 meV smaller than expected because the authors don't directly measure the $E_{1s}$ via absorption spectroscopy, but rather, infer the levels by adding a 50 meV Stokes shift to the PL emission line. Nevertheless, Chen et al. demonstrated that $E_g$ for 2D HPs can be probed via 2P-PLE and may have resolved faint signatures of the $2p$ states as well. Incidentally, these step-like absorption features marking $E_g$ are even more clear in single-photon PLE experiments.[155] In the case of PEA$_2$PbI$_4$, the feature can even be resolved in high-quality samples at low temperature using single-photon transmission or reflection spectroscopies [19, 124], however, for BA$_2$PbI$_4$ the proximity of an above-gap exciton peak prevents its detection in this manner.

In summary, resolving $n > 1$ members of the exciton's Rydberg series in some cases is easier than measuring $E_g$. This is most often the case for low-dimensional semiconductors, where the interband oscillator strength is weak due to the low dimensionality, but exciton absorption is enhanced. For TMDC monolayers and 2D HPs, optical signatures of the $n > 1$ Rydberg peaks are faint even in high-quality samples at low temperature. Yet, extremely clear Rydberg features have been observed using magneto-optical techniques on TMDC monolayers encapsulated in hBN (see **section 7.1**). While many studies have focused on the *energy* ladder of Rydberg states in low-dimensional materials,[66] the *oscillator strengths* as a function of $n$ have not been studied as deeply.

The observed ratio of the oscillator strengths between 1s:2s seems to vary quite drastically for different materials. For examples, the optical signatures assigned to the $2s$ state in TMDC monolayers and 2D HPs are typically 1% to 2% the magnitude of the $1s$ state, [64, 66, 209] much lower than the 3D hydrogen model which predicts a proportionality of ~12% and the 2D hydrogen model that predicts ~4%. [17] Future experimental and theoretical investigations along these lines would be nice complements to the multiple studies focused on the non-hydrogenic energy levels.

## 6. Electroabsorption spectroscopy

Electroabsorption spectroscopy (EA) measures electric field-induced shifts in a material's absorption spectrum, and is defined as $\Delta A = A(F) - A(F = 0)$, where $F$ is the electric field. Generally speaking, modulation spectroscopy techniques such as EA are superior to linear absorption as a probe of critical points in a semiconductor's joint-density of states, such as the $E_{1s}$ or $E_g$ energy levels. This is because EA is measured via lock-in modulation as a normalized difference signal, either in transmission $\Delta T/T$ or reflection $\Delta R/R$ geometries, where background artifacts such as optical scattering are eliminated and the widths of the spectral features are determined purely by lifetime broadening. [142, 212] So long as the EA device has no current leakage or Joule heating, then one can be quite confident that the EA lineshape is intrinsic to the properties of the material itself. This lineshape is rich with information and if the proper conditions are met, can yield trustworthy measurements of the exciton's dipole moment ($\mu$), polarizability ($\alpha$), radius ($r$), binding energy ($E_b$), and mass ($m^*$). [124] To measure $E_b$ using EA, the measurement strategy differs depending on the magnitude of $E_b/\Gamma$; from our experience performing electroabsorption on 2D and 3D HPs, the boundary is roughly $E_b/\Gamma \sim 2$. In the following, we discuss the differences between the two regimes and show examples of successful $E_b$ measurements in both the large $E_b$ regime ($E_b/\Gamma > 2$) and the low $E_b$ regime ($E_b/\Gamma < 2$).

### 6.1 $E_b/\Gamma > 2$ regime

In the case of relatively large $E_b$ and low homogeneous broadening $\Gamma$, the exciton and interband features have sufficient spectral separation for the features to be resolved independent of one another. Furthermore, since electric fields affect exciton and continuum states differently, the field-dependence of the EA amplitude and lineshape often reveals the nature of the underlying EA feature which can allow for unambiguous assignments of the respective features.

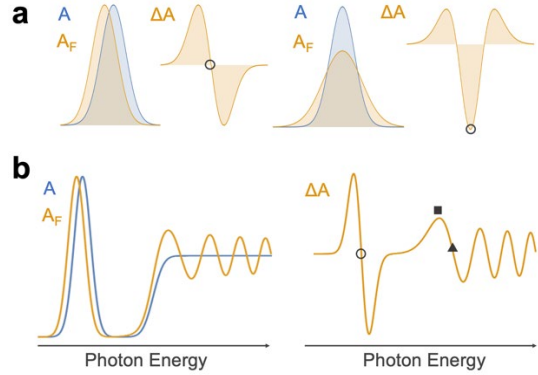

**Figure 13**: **(a)** The shifting of an absorption peak leads to a first derivative lineshape in the EA signal (left) whereas the broadening of the peak leads to a second-derivative lineshape (right). Here, the $E_{1s}$ energy is indicated by the open circle. **(b)** Schematic representing the standard EA response of a semiconductor with polarizable (Wannier) excitons. Below the band-gap, the exciting peak redshifts due to the quadratic stark effect. Oscillations occur at the interband transition according to the Franz-Keldysh effect. In a 3D semiconductor, $E_g$ occurs at the first Franz-Keldysh peak (black square) whereas in a 2D semiconductor $E_g$ is marked by the first zero-crossing (black triangle).

Below the band-gap, the exciton peak will redshift proportional to the quadratic Stark effect and broaden proportional to the linear Stark effect. As depicted in **Figure 13a**, the shifting and broadening of a Lorentzian or Gaussian-shaped peak produces first and second derivatives in the lineshape of the difference signal (EA). As such, the EA signal can be represented by the following Taylor expansion: [213]

$$\Delta A = \frac{\partial A}{\partial E}\Delta E_Q + \frac{1}{2}\frac{\partial^2 A}{\partial E^2}(\Delta E_L)^2 \qquad (19)$$

Here, $\Delta E_L$ and $\Delta E_Q$ and represent the exciton's energy shift resulting from the linear and quadratic Stark effects. Their magnitudes are proportional the exciton's dipole moment and polarizability, as follows:

$$\Delta E_L = -\mu_{ge}F \qquad (20)$$

$$\Delta E_Q = -\frac{1}{2}\alpha_{ge}F^2 \qquad (21)$$

where $\mu_{ge}$ and $\alpha_{ge}$ are the change in permanent dipole moment and polarizability, respectively, upon excitation from the ground to excited state $g \to e$. Thus, if the exciton's absorption and EA features are resolved and the magnitude of $F$ within the EA device is known, then $\mu_{ge}$ and $\alpha_{ge}$ can be accurately measured. An ideal Wannier exciton has no permanent dipole moment since 1s wavefunction is spherically symmetric. Thus, for Wannier excitons, $\Delta E_Q \gg \Delta E_L$ and the EA lineshape is primarily first-derivative. In this case the $E_{1s}$ position is most accurately determined by zero-crossing in the EA signal. However, if the EA lineshape has a significant second derivative component then it's generally more accurate to take the exciton's absorption peak as $E_{1s}$.

Continuum states respond to applied electric fields very differently, through the Franz-Keldysh (FK) effect. The addition of an electric field potential, $eFz$, to the free electron's Hamiltonian results turns the free electron wavefunction from a Bloch function to an Airy function and it has been shown that the semiconductor's absorption also acquires an Airy-like behavior at high field strengths: [214]

$$\Delta A \propto Ai\left(\frac{E-E_g}{\hbar\theta}\right) \qquad (22)$$

where $\hbar\theta$ is the electro-optic energy, defined as: [214]

$$\hbar\theta = \left(\frac{\hbar^2 e^2 F^2}{2m^*}\right)^{1/3} \qquad (23)$$

According to the form of the Airy function, the amplitude of FK features *broaden* in energy proportional to $F^{2/3}$ and increase in amplitude proportional to $F^{1/3}$. This is contrast to Stark effects on bound states which produces *field-invariant* EA lineshapes with amplitudes that increase proportional to $F^2$, according to **Equation (19)**. Thus, the field-dependence of the EA signal can be a key factor allowing for clear discernment between exciton and interband features. According to Airy function in **eq (22)**, the FK effect influences interband absorption in two major ways: first, there is a transfer of oscillator strength from above to below the band gap and second, there are Airy-like oscillations above the bandgap (see **Figure 13b**).

The first of the two effects produces a single oscillatory period in the EA spectrum. For a 3D semiconductor, it has been shown that the first peak in this oscillation (marked by the square in **Figure 13b**) is very close to the one-electron band gap energy $E_g$. [215] For a 2D semiconductor is was recently shown that $E_g$ is more accurately determined by the zero-crossing (marked by the triangle in **Figure 13b**). [124] Thus, in the case that both Stark and FK features are independently resolved, $E_b$ can be determined in a trustworthy fashion by $E_b = E_g - E_{1s}$. In the following, we showcase two examples from the literature where $E_b$ was measured in this manner. In each case, the field-dependence of the EA lineshape and amplitude allowed the exciton and continuum features to be distinguished from one another.

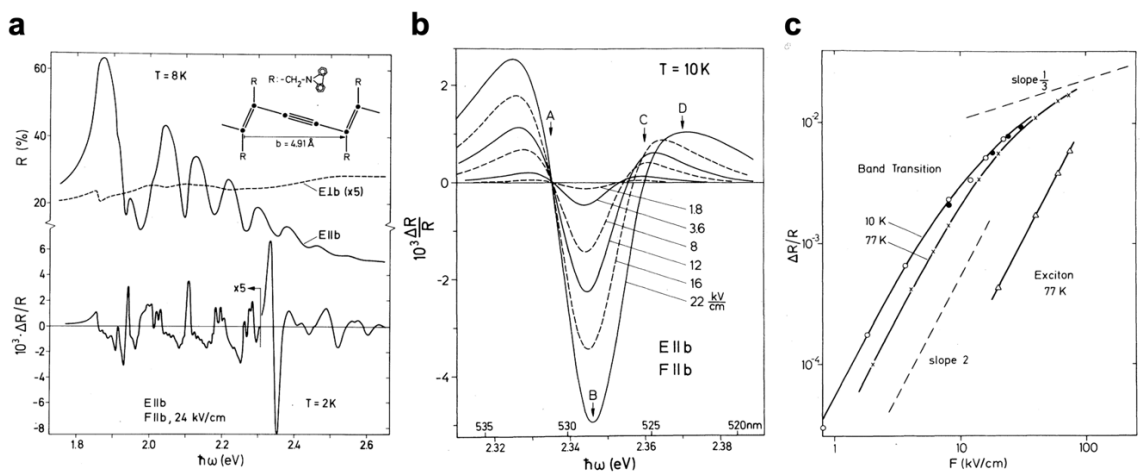

**Figure 14**: **(a)** (Top) Reflection spectrum of a polydiacetylene single crystal measured with light polarized perpendicular to the polymer chain (E⊥b) and parallel to the polymer chain (E//b). (Bottom) ER signal where both the applied field and the photon polarization are parallel to the polymer chain. **(b)** The same ER signal with a zoomed-in horizontal axis near the band gap ($2.31 < \hbar\omega < 2.39$ eV), plotted for a range of electric field strengths. **(c)** The amplitude of the ER signal as a function of electric field strength. Panels (a-c) are reprinted from ref. [176] with permission from the American Physical Society, Copyright 1981.

The first clear measurements of $E_b$ in an organic semiconductor were obtained by L. Sebastian and G. Weiser in 1981. [176] In their study, metal contacts were evaporated onto the surface of a polydiacetylene single crystal (molecular structure shown in **Figure 14a**) and electroreflection (ER) measurements were taken at a near-incident angle. **Figure 14a** displays the polydiacetylene reflection spectra for photon polarizations parallel (solid black) and perpendicular (dotted black) to the polymer chain. The first indication of the one-dimensional nature of the polydiacetylene electronic structure comes from the observation that the exciton signatures in the reflection spectrum are about two orders of magnitude greater in the parallel geometry. Additionally, as is clear from the ER signal at the bottom of **Figure 14a**, these exciton features are sensitive to electric fields. The first-derivative lineshape and the large magnitude of the exciton's ER signal ($\Delta R/R \sim 10^{-4}$ at a modest field strength of 24 kV/cm) suggests that these excitons polarizabilities similar to those in inorganic semiconductors and therefore, are Wannier in nature. [177]

The major discovery in their study, however, is a weaker signal near 2.3 eV with a field-dependence expected for FK features. In **Figure 14b**, this signal is plotted for several electric field strengths. Between the points labeled A and D, the ER lineshape broadens with increasing field and the amplitude increases sub-quadratically. This field dependence is shown in **Figure 14c**, where the amplitude of the exciton and band-edge signals are plotted side-by-side as a function of applied field. While the exciton has consistent slope of 2 across the entire field-range, the band-edge has a slope that tapers off at higher fields, approaching a slope of 1/3 at the highest field strengths (~70 kV/cm) which is expected for the high-field FK effect. Accordingly, the authors chose the onset of this feature (point A) as $E_g$, resulting in an exciton binding energy of 480 meV. While their selection of $E_g$ *within* the FK feature is not discussed (the FK effect in a one-dimensional system is not well understood), the error on their assignment of $E_g$ is only ~ 20 meV, as the large field-broadening near points 'C' and 'D' clearly indicate these features lie above the band gap. In our estimation, this measurement of $E_b$ by Sebastian et al. is probably the most trustworthy and precise measurement of $E_b$ in an organic semiconductor to-date because optical signatures of $E_g$ are so rare for organic systems. One's confidence in their assignment of the FK feature should only increase in light of recent EA measurements by Hansen et al., where a nearly identical signal was resolved at the band gap of a 2D HP thin film (**Figure 15**). [124]

    The full EA spectrum for the 2D HP thin film, with a composition of PEA$_2$PbI$_4$, is shown in **Figure 15a**. The features labeled $a - c$ represent the Stark shift of the exciton while those labeled $d - h$ are a result of the FK effect, similar to the feature resolved in polydiacetylene. The amplitude and shape of this FK feature exhibit the same field dependence observed for polydiacetylene, but there are two additional pieces of evidence in the 2D HP case that further confirm its physical origin. First, as shown in **Figure 15a**, the oscillation in the EA spectrum coincides with a step-like function in the linear absorption, indicating its origin lies in an interband transition. Second, a nearly identical signal is observed at the band gap of a simulated EA spectrum for a 2D Wannier exciton. The zoomed-in version of this FK signal, for a butylammionuum lead iodide (BA$_2$PbI$_4$) thin film, is displayed in **Figure 15b**. As discussed in their study, the points $d - h$ have a one-to-one correspondence with an oscillation near $E = E_g$ in the simulated EA spectrum. [124]

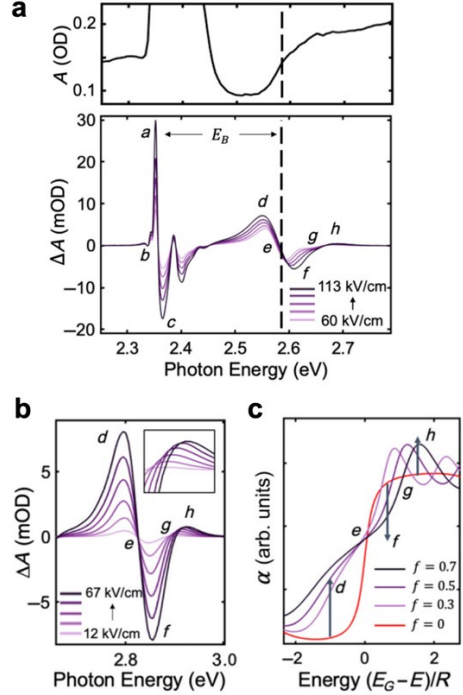

**Figure 15**: **(a)** Absorption (top) and electroabsorption (bottom) of a PEA$_2$PbI$_4$ thin film. **(b)** Zoom in on the FK feature. **(c)** Simulated zero-field (red) and field-perturbed (purple) absorption spectra for a 2D Wannier exciton. Panels (a-c) are reprinted from ref. [124] with permission from the American Physical Society, Copyright 2022

Here, the simulated EA is produced by solving the 2D Schrödinger equation for an electron-hole pair in a uniform field and relating the resulting wavefunctions to the material's absorption through Elliot's formula. [17, 216] When the simulated zero-field (red) and field-shifted (purple) absorption spectra are plotted side-by-side (see **Figure 15c**), it is clear that the positive and negative differences underlying this FK feature result from a transfer of oscillator strength across the band gap. In accordance with the form of the conduction band's Airy function, the continuum wavefunction leaks further into the forbidden gap as the field strength increases thereby increasing the absorption below $E_g$ and decreasing the absorption above $E_g$. The simulation shows that the cross-over point (zero crossing in the EA signal) marks the band gap energy. The authors took the zero-crossing of the exciton's first derivative line shape as $E_{1s}$, and found $E_b = 223 \pm 3$ meV for PEA$_2$PbI$_4$ and $E_b = 250 \pm 4$ meV for BA$_2$PbI$_4$.

## 6.2 $E_b/\Gamma < 2$ regime

In the above EA studies, the measurements were taken at low temperature where $\Gamma$ is low and using a high-$E_b$ material. Thus, the exciton and FK feature were spectrally separated and the condition of $E_b/\Gamma > 2$ was well satisfied. However, these conditions are not always experimentally accessible. For example, in a recent study by Ziffer et al, the technologically relevant 3D HP material MAPbI3 was studied at room temperature. [33] Cooling the sample to reduce $\Gamma$ was not an option since MAPbI3 crystal undergoes a phase transition at low temperatures, and the authors wished to measure $E_b$ under solar cell operating conditions in order to shed light on whether perovskite solar cells are free-carrier or excitonic in nature. This is a particularly arduous task given that the binding energy relative to the broadening is tiny for MAPbI3 at room temperature, with $E_b/\Gamma \sim 0.29$ ($\Gamma \sim$ 35 meV, $E_b \sim$ 10 meV). Under these conditions, bound exciton states and continuum states are effectively "mixed" because the excitons are easily ionized at zero-field by scattering events with phonons or charged impurities. [217, 218] If the condition of $\hbar\theta < \Gamma/3$ is met (and it nearly always will be if $E_b/\Gamma < 2$), then the material falls into the low-field FK regime and the EA response is proportional to the third derivative of the complex dielectric function ($\varepsilon = \varepsilon_r + i\varepsilon_i$), as follows: [142, 219]

$$\Delta\varepsilon(E,F) = \frac{(\hbar\theta)^3}{3E^2}\frac{d^3}{dE^3}(E^2\varepsilon(E,0)) \tag{24}$$

where $E$ is the photon energy and $\Delta\varepsilon$ represents the field-induced changes to both the real and imaginary dielectric function: $\Delta\varepsilon_r + i\Delta\varepsilon_i$.

The strategy for measuring $E_b$ under these conditions was pioneered by Ziffer et al., and it is as follows: (1) Use Elliot's formula to describe $\varepsilon(E, 0)$ in terms of parameters $E_b$, $E_g$, and $\Gamma$ (these are the fit parameters). (2) Convert $\varepsilon(E, 0)$ to $\Delta\varepsilon_r$ and $\Delta\varepsilon_i$ using **Equation (24)**, then convert $\Delta\varepsilon_r$ and $\Delta\varepsilon_i$ to $\Delta n$ and $\Delta\kappa$ as per their standard definitions. (3) Compute the field-shifted optical constants ($n_F$, $\kappa_F$) using ($n + \Delta n$, $\kappa + \Delta\kappa$); here, $n$ and $\kappa$ must be obtained from independent ellipsometry measurements while $\Delta n$ and $\Delta\kappa$ are known from step #3. (4) Use transfer matrix modeling to compute $T_F$ from ($n_F$, $\kappa_F$) and $T_0$ from ($n, \kappa$); the simulated EA spectrum is then $(T_F - T_0)/T_0$. (5) Finally, using a least-squares regression method, fit the simulated $\Delta T/T$ to the experimental $\Delta T/T$ while adjusting the fit parameters $E_b$, $E_g$, and $\Gamma$. While this is a non-trivial fit procedure and requires additional ellipsometry measurements, the reward is measurement of $E_b$ and $E_g$ within a regime where other measurement techniques may not be possible.

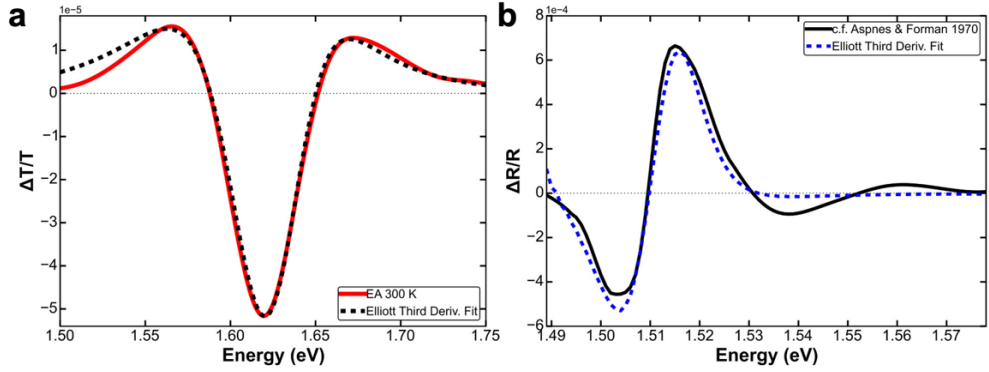

**Figure 16**: **(a)** The band edge electroabsorption of MAPbI$_3$ (red) and the results of the third-derivative Elliot fit (dotted black line). The fit procedure is described in the main text. **(b)** The band edge electroreflectance of GaAs (black) and the corresponding third-derivative Elliot fit (dotted blue). Panels (a-b) are reprinted from ref. [33] with permission from the American Chemical Society, Copyright 2016.

In their study, Ziffer et al. applied this procedure to MAPbI$_3$ and found good agreement between the experimental and simulated EA lineshapes, as shown in **Figure 16a**. The least squares regression produced an $E_b$ value of 7.4 meV (95% confidence interval spanning 6.8 – 9.2 meV) which is in good agreement with high-field magnetoabsorption measurements on the same material. [101] To test the validity of the measurement, the fit procedure was also applied to GaAs (**Figure 16b**) where a value of 3.8 meV was obtained, close to the value of 4.2 meV obtained from resolving the 2s exciton peak via low-temperature absorption.

## 7. Magneto-optical spectroscopy

Magneto-optical spectroscopy measures magnetic field-induced shifts in a material's transmission, reflection, absorption or PL spectrum and is among the most popular techniques to probe an exciton's properties. In the low magnetic field regime, the exciton peak energy shifts according to the diamagnetic coefficient, which can be measured and converted into the exciton's radius, mass, and binding energy. However, the conversion is sensitive to the choice of the dielectric constant. Binding energy measurements within the high-field regime are more desirable. The high-field condition of $\hbar Be/m^* > E_b$ (where $B$ is the magnetic field strength) is easily met for III-V systems due to their low carrier masses and low exciton binding energies, but is experimentally challenging for HPs (requiring $B > 50$ T), and experimentally off-limits for TMDCs due to their high-carrier mass and high exciton binding energies. In the following, we provide a brief background on magneto-optics as a probe of $E_b$ in both regimes and highlight important measurements from the literature.

**7.1 Low-field regime:**

When the energy related to the cyclotron motion of free carriers ($\hbar Be/m^*$) is much lower than $E_b$, the magnetic field-induced shift of the $1s$ exciton peak be described as follows: [220]

$$\Delta E(B) = \pm \frac{g\mu_B}{2}B + c_0 B^2 \qquad (25)$$

Here, $g$ is the Landé g-factor, $\mu_B$ is the Bohr magneton, and $c_0$ is the diamagnetic coefficient. The first term in **Equation (25)** is from the Zeeman effect while the second term represents the exciton's diamagnetic shift. As graphically depicted in **Figure 17a**, the Zeeman effect splits the conduction and valence bands into nondegenerate spin-split bands. The figure depicts this for J=1/2 conduction and valence bands, but the details can be difference for different materials. For example III-V semiconductors typically have a J=3/2 valence band as a consequence of the combination of S=1/2 for electrons with the L=1 character of the orbitals comprising the VB. [221] The absorption spectra for left-hand ($\sigma^+$) and right-hand ($\sigma^-$) circular polarizations will be shifted in energy according to spin-conserving selection rules for optical transitions. Assuming the exciton in Wannier in nature, the measured diamagnetic coefficient can be used to estimate $E_b$ as follows: [116, 117]

$$\frac{E_b}{R_H} = \left(\frac{1}{\varepsilon_r^4}\frac{c_H}{c_0}\right)^{1/3} \qquad (26)$$

where $c_0$ is the diamagnetic coefficient for hydrogen. Unfortunately, this route to obtaining $E_b$ has the drawback that, according to **Equation (26)**, it is more sensitive $\varepsilon_r$ as an input parameter than the experimentally determined $c_0$. The uncertainty in $\varepsilon_r$, which is typically large both due to measurement and ignorance surrounding the correct frequency value to select, propagates to considerable uncertainty to $E_b$. [114]

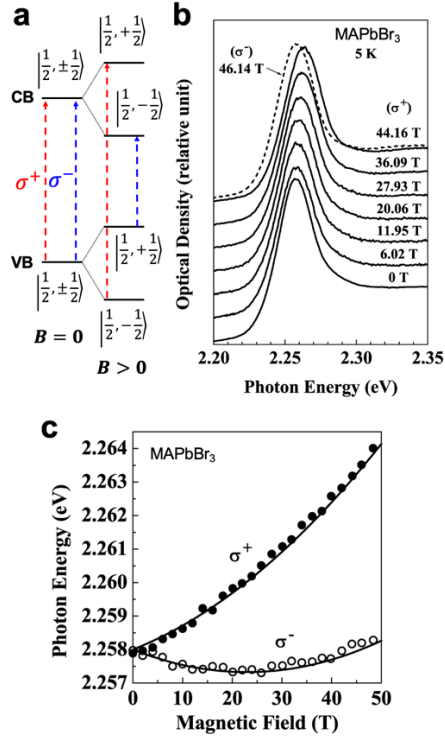

**Figure 17: (a)** Under a magnetic field, the spin-split CB and VB will absorb $\sigma^+$ and $\sigma^-$ polarizations of light differently. **(b)** Absorption spectra for MAPbBr$_3$ under various magnetic field strengths, where the solid line represents $\sigma^+$ absorption and the dotted line represents $\sigma^-$ absorption. **(c)** The exciton resonance energy as a function of magnetic field for the $\sigma^+$ and $\sigma^-$ polarizations, fit to **Equation (25)**. Panels (b-c) are reprinted from ref. [117] with permission from Elsevier, Copyright 2003.

Throughout the 1990s and early 2000s, this strategy was used to estimate $E_b$ for 2D HPs [222] and 3D HPs. [116, 117] For example, in their 2003 study, Tanaka et al. measured the magnetoabsorption spectra up to $B = 45$ T for 3D HP compositions with iodide and bromide. [117] The absorption spectra for methylammonium lead bromide (MAPbBr$_3$) under increasing magnetic field strengths are shown **Figure 17b**, where the solid lines correspond to absorption of $\sigma^+$ polarized light and the dotted light is for $\sigma^-$ polarization. At a magnetic field of 44.16 T, the splitting of the exciton peak by ~10 meV is visible to eye. This relatively large shift allows for accurate tracking of the peak energy as a function of magnetic field, and these $\Delta E(B)$ values are plotted in **Figure 17c**. The solid line represents **Equation (25)** with the fitted values of $g = 2.03$ and $c_0 = 1.28 \times 10^{-6}$ eV/T$^2$. While this measurement of $c_0$ is trustworthy and robust, the subsequent conversion to $E_b$ according to **Equation (26)** is extremely sensitive to the choice of $\varepsilon_r$. Tanaka et al. opted to use the optical-frequency value, $\varepsilon_r = 4.8$, resulting in $E_b = 76$ meV. However, it has since been shown through DFT calculations that $\varepsilon_r$ at infrared frequencies, not optical frequencies, determines $E_b$ in HPs. [109] Accordingly, the Coulomb interaction is well screened and $E_b$ is quite low— state-of-the-art high-field magnetoabsorption indicate values of 25 meV for MAPbBr$_3$ [100] and 10 meV for MAPbI$_3$ [101] (discussed more below).

While converting $c_0$ to $E_b$ has been less popular within the literature on III-V semiconductors and TMDCs, measurements of $c_0$ have been important for determining the exciton's radius [74, 78] and reduced effective mass in these systems [51, 223], as the radius and mass depend on $c_0$ in a manner that is less sensitive to $\varepsilon_r$. [116, 117] For magnetoabsorption measurements on III-V systems, $E_b$ is reliably obtained in the high-field regime, as discussed further below. TMDCs, on the other hand, have high free carrier masses (~0.5 m$_0$ [60]) and exciton binding energies (> 100 meV) mandating unreasonably large magnetic fields of $B > 300$ T to fulfill the high-field condition. Nonetheless, some of the highest quality $E_b$ measurements on TMDC monolayers resulted from magneto-optical experiments in the low-field regime due to a clear signal originating from the diamagnetic shift of the exciton's Rydberg series.

The diamagnetic shift of the $ns$ Ryberg series in a TMDC monolayer was first reported by Steir, Crooker, and colleagues in 2018. [76] Previously, the diamagnetic shift of the $1s$ state for TMDC monolayers on SiO$_2$ were studied by the same authors in 2016, leading to important measurements of the exciton's radius $r$ and theoretical calculations of $E_b$ based on the Rytova-Keldysh potential (**Equation (6)**). [77, 78] However, in these earlier studies the doping caused by the SiO$_2$ substrate led to broad absorption features and the Rydberg series was not resolved.

To reduce Γ in their subsequent 2018 study, the WSe$_2$ monolayer was enclosed in hBN and mounted to the tip of an optical fiber inside of a pulsed magnet, as depicted in **Figure 18a**. The suppressed doping as a result of the hBN encapsulation reduced Γ from ~25 meV to ~9 meV, and as a result, the exciton's Rydberg series up to the $4s$ state was resolved. The magneto-transmission spectrum of the encapsulated WSe$_2$ monolayer is shown in **Figure 18b** for both $\sigma^+$ and $\sigma^-$ configurations. The $2s$ feature near 1.85 eV is clearly resolved and its energy shifts with magnetic field as expected for a feature with excitonic origin. When the transmission spectrum is plotted as 2D image plot as a function of magnetic field (**Figure 18c**), the $3s$ and $4s$ states are also resolved and especially clear at high magnetic fields.

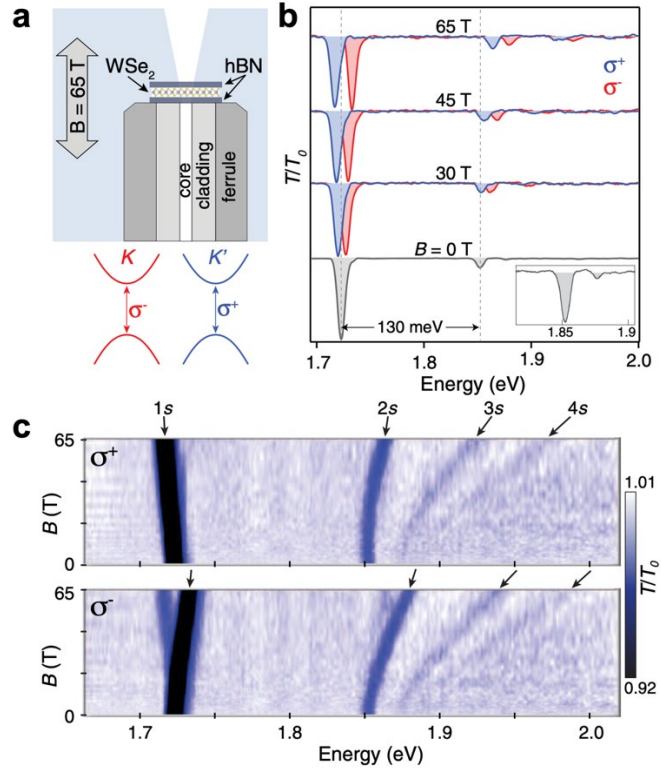

**Figure 18**: **(a)** (top) A common experimental set-up for measuring the diamagnetic shifts of TMDC monolayers. The monolayer is mounted to the tip of an optical fiber and mounted inside of a cryostat and pulsed magnet. In this case, the monolayer is encapsulated in hBN to reduce substrate doping effects. (bottom) Due to broken inversion symmetry, the Brillouin zone of TMDC monolayers hosts distinct K and K' points with opposite spin-orbit splitting and can be selectively probed using $\sigma^-$ and $\sigma^+$ polarizations, respectively, leading to a valley Zeeman effect, as described in Ref. [78]. **(b)** Left-hand $\sigma^+$ (blue) and right-hand $\sigma^-$ (red) polarized transmission spectra of the WSe$_2$ monolayer under various magnetic field strengths. **(c)** The same data in panel (b), but the transmission is plotted as a 2D image with greater resolution in magnetic field. Panels (a-c) are reprinted from ref. [76] with permission from the American Physical Society, Copyright 2018.

Here, the energy shifts are caused by both Zeeman and diamagnetic shifts, although the physics underlying the Zeeman component in monolayer TMDCs is slightly different from the traditional Zeeman effect described previously due to distinct points in the Brillouin zone, termed $K$ and $K'$, with equal-but-opposite magnetic moments (see **ref.** [78] for details on this point). Nevertheless, Stier et al. are able to measure the $\Delta E(B)$ for $n = 1s, 2s, 3s,$ and $4s$ exciton states, subtract the Zeeman effect by averaging $\Delta E(B)$ for $\sigma^+$ and $\sigma^-$ polarizations, and convert the measured diamagnetic coefficient into measurements of the exciton's radius ($c_0 \propto \langle r^2 \rangle$). The $3s$ and $4s$ diamagnetic shifts are particularly interesting as they satisfy a strong diamagnetic field limit of $\langle r^2 \rangle > \sqrt{\hbar/eB}$, where the slope and separation of these states can be used to determine the exciton's reduced effective mass [60, 76]. The authors find $m^* = 0.20 \pm 0.1$ m$_0$ for WSe$_2$ and published another study the following year where this method of measuring $m^*$ is extended to MoS$_2$, MoSe$_2$, and WS$_2$. [60] In terms of measuring $E_b$, the clearly resolved Rydberg series up to $n = 4$ makes for an extremely trustworthy measurement of $E_b = 161$ meV for WSe$_2$ [76] ($E_b = 167$ meV was reported the following year by the same group [60]). A study by Molas et al. in 2019 found an equally clear signal up to $n = 5$ in the magneto-PL spectrum of WSe$_2$ and also found $E_b = 167$ meV. [61] The $E_b$ values for the other TMDC compositions are also in excellent agreement between the two groups. [60, 61]

Incidentally, both studies (Steir et al. [76] and Molas et al. [61]) report that their hBN-encapsulated TMDC monolayers with $E_b \sim 200$ meV exhibit a Rydberg energy ladder that is far closer agreement with the hydrogenic model than previously reported for non-encapsulated TMDC monolayers with $E_b \sim 400$ meV. As described in **section 5**, these earlier observations of a non-hydrogenic Rydberg series by He et al. [64] and Chernikov et al. [66] were based on faint, yet reproducible features in the reflectance spectrum of TMDC monolayers between SiO$_2$ and vacuum. Steir et al. argue this disparity in the observed Rydberg energy ladder is merely a consequence of the large hBN dielectric constant and the authors show that the Rytova-Keldysh potential (**Equation (6)**) predicts a nonhydrogenic series for low-dielectric environments and a hydrogenic series for high-dielectric environments ($\varepsilon_t = \varepsilon_b = 4.5$ for hBN encapsulation).

Furthermore, Steir et al. report that their measured exciton radii deviate from the hydrogenic model in a manner that is consistent with previous reflectance studies. Molas et al., on the other hand, chose to fit their Rydberg energies to a hydrogenic model and propose that a different potential (Krazer potential) more accurately describes excitons in TMDC monolayers. [61] Interestingly, in this same year a study by Chen et al. measured both the magneto-PL and reflectance spectra for the same hBN-encapsulated WSe$_2$ monolayer. [92] The faint, Rydberg-like features in the reflectance were found to be energetically aligned with the clear Rydberg features in the magneto-PL, thereby validating the previous assignments and non-hydrogenic energy ladder reported by He et al. and Chernikov et al. in their 2014 reflectance studies. Thus, a consistent picture has emerged where excitons in TMDC monolayers are well-described by the Rytova-Keldysh potential, with a hydrogenic Rydberg series and $E_b \sim 200$ meV under hBN encapsulation, but a non-hydrogenic Rydberg series and $E_b \sim 400$ meV for monolayers between SiO$_2$ and vacuum.

These magneto-optical $E_b$ measurements by Steir et al., [60, 76] Molas et al., [61] and Chen et al. [92] are probably the most trustworthy and precise $E_b$ measurements for monolayer TMDCs due to the clarity and the field-dependence of the signal up to $n = 4$ and $n = 5$. Thus, magneto-optical techniques have been critically important for studying excitons in TMDC monolayers despite the fact that TMDCs fall into the low magnetic field regime. Interestingly, TMDC monolayers can be artificially pushed into the high-field regime via extremely high doping concentrations, and in these cases, transitions between Landau levels have been experimentally observed. [224-226] However, these experiments are beyond the scope of this review as they generally do not produce $E_b$ measurements.

### 7.2 High-field regime:

The high-field regime is entered when the energy associated with the cyclotron motion of free electrons and free holes is greater than the exciton binding energy ($\hbar Be/m^* > E_b$), or equivalently when the cyclotron gyroradius is greater than the exciton's radius. [220] Under these conditions, a series of above-gap absorption peaks can appear due to transitions between van-Hove singularities of the Landau levels. The energies at which these features appear is given by: [101, 220]

$$E_N = E_g + \left(N + \frac{1}{2}\right)\left(\frac{\hbar e}{m^*}\right)B \pm \frac{1}{2}g_{eff}\mu_B B \qquad (27)$$

where $N = 0, 1, 2, 3$, etc. is the orbital quantum number of the Landau level and $g_{eff}$ is the effective $g$-factor. If unpolarized light in used, then the last term can be ignored and the peaks of the Landau transitions $E_N$ can simply be fit to a line according to the first term in **Equation (27)**, where the y-intercept is $E_g$ and the slope gives the mass of the exciton's reduced mass $m^*$. The exciton binding energy can then be straightforwardly calculated in a model-independent fashion by $E_b = E_g - E_{1s}$, where the 1s position is easily obtained from the exciton absorption peak at zero magnetic field. This has been a very successfully strategy for measuring $E_b$ in III-V semiconductors and HPs.

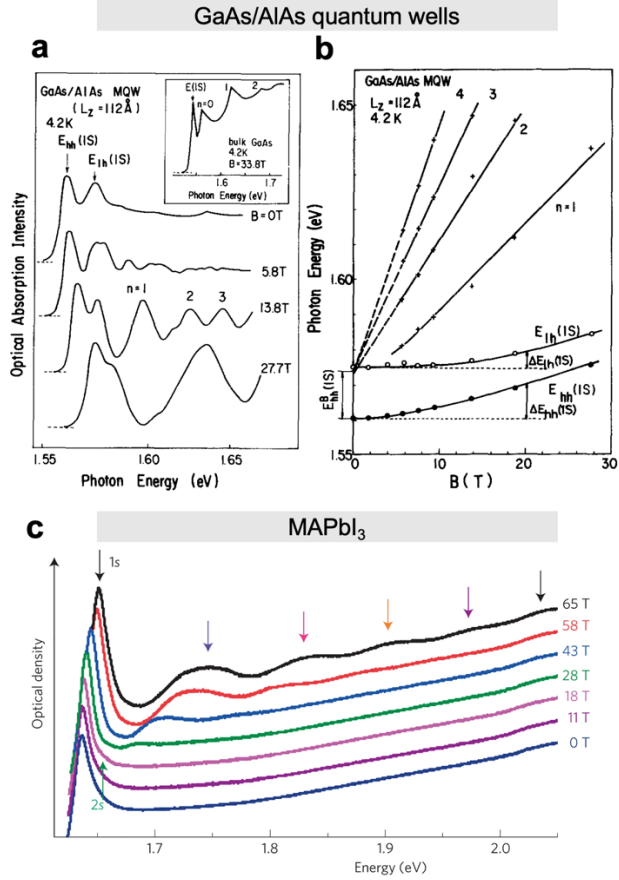

**Figure 19**: **(a)** Absorption spectra of a GaAs/AlAs quantum well structure at magnetic field strengths of $B = 0$, 5.8, 13.8, and 27.7 T. The absorption peaks associated with transitions between Landau levels are label 1, 2, and 3. **(b)** The peak energies as a function of magnetic field. The energy of the heavy-hole (hh) and light-hole (lh) exciton peaks are represented by closed and open circles, respectively, while the $N = 1, 2, 3, 4$ Landau level associated peaks are marked by '+' symbols. Panels (a-b) are reprinted from ref. [31] with permission from Elsevier, Copyright 1984. **(c)** The absorption spectrum of MAPbI$_3$ at magnetic field strengths ranging from 0 to 65 T. The regularly-spaced peaks originating from interband Landau level transitions become very evident at $B > 50$ T. Reprinted from ref. [101] with permission from Springer Nature, Copyright 2014.

Due to the low $E_b$ and $m^*$ values in III-V semiconductors, the high-field condition can be met at relatively modest magnetic field strengths, as shown in **Figure 19a** where the Landau level associated absorption peaks for a GaAs/AlAs quantum well structure are evident at $B = 5.8$ T and become progressively large with increasing field strength (these data were collected by Tarucha et al. [31]). The $N = 1, 2, 3$, etc transitions between Landau levels are labeled in **Figure 19a** and the peak energies are plotted as a function of magnetic field in **Figure 19b**. The higher-transitions have greater slopes, according to **Equation (27)** where the slope is proportional to $N + 1/2$, and the lines clearly converge to a common y-axis intercept resulting in a precise measurement of $E_g = 1.573$ eV and a heavy-hole exciton binding energy of ~12 meV. The data shown here is for a device with a well length of 112 Å. Tarucha et al. repeat these magneto-optical measurements for a range of quantum well lengths in order to determine the relationship between $E_b$ and well length. Throughout the 1980s and 1990s, magnetoabsorption studies such as this were widespread and very important in assessing the validity of competing theoretical models for excitons in quantum wells. [30-32, 44]

This strategy has been applied in an identical manner to measure $E_b$ and $m^*$ in 3D HPs [100, 101, 144] and 2D HPs. [103, 125] Magnetoabsorption on HPs in the high-field regime was first reported by Miyata et al. in 2015. The central result from their study is shown in **Figure 19c**, where the absorption spectrum of MAPbI$_3$ is plotted as a function of magnetic field. The peaks associated with transitions between Landau levels rapidly become increasing clear for $B > 50$ T, meaning that the previous low-field measurements by Tanaka et al. from 2005 were only short ~10 T from entering the high-field regime and obtaining much better $E_b$ measurements.

In their study, Miyata et al. show that each of these interband Landau peaks has a magnetic field dependence that is expected from **Equation (27)**. The peak energies $E_N(B)$ are linear with field and tightly converge to a common y-intercept of $E_g = 1.652$ eV, resulting in $E_b = 16 \pm 2$ meV. The slope yielded a reduced effective mass of $m^* = 0.104 \pm 0.003 m_0$. Miyata et al.'s measurement was reproduced in 2017 by Yang et al. using high-field magnetoreflectance on an MAPbI$_3$ single crystal. [150] However, a recent 2021 study by Yamada et al. obtained a much lower value of $E_b = 3.1 \pm 0.5$ meV by assigning Landau transitions at low magnetic fields. [151] The authors comment that this discrepancy (3.1 vs 16 meV) is due to the difficulty in extrapolating the high-field energy levels down to B = 0, and we recommend the data analysis to be reattempted by outside groups to assess this claim. Nevertheless, these inter-Landau transitions are generally considered to yield the most trustworthy measurements of $E_g$, $E_b$ and $m^*$ in 3D HPs. Many other $E_b$ measurement methods fail when applied to 3D HPs since $E_b$ and $\Gamma$ are of the same magnitude even at low temperatures. The inter-Landau transitions, however, are only resolved at low temperatures and this leaves a slight knowledge gap for the temperature dependence of $E_b$. Although, it is generally accepted, from electroabsorption measurements [33] and DFT calculations, [109] that $E_b$ is reduced slightly in the room temperature tetragonal phase which is more relevant to the performance of MAPbI3 in a solar device. Similar high-field magneto-optical studies have been conducted on 2D HPs [103, 125] and the binding energies are in general agreement with electroabsorption studies. [115, 124]

## 8. Scanning Tunneling Spectroscopy

As is evident from the above studies, optical detection of the fundamental band gap ($E_g$) is challenging and is most likely impossible for some systems such as TMDC monolayers and many organic semiconductors. In these systems in particular, the interband absorption is weak and is obfuscated by neighboring exciton peaks. When traditional optical methods fail, there are two common alternative methods to obtain $E_g$, namely scanning tunneling spectroscopy (STS) or a combination of inverse photoelectron spectroscopy (IPES) and ultraviolet photoelectron spectroscopy (UPS). There are many parallels between STS and IPES/UPS— both techniques are surface sensitive, probe transitions of real electron populations, have relatively broad spectral features in comparison to optical spectroscopies, and lastly, both techniques require similar fitting procedures to extract $E_g$. Because STS is more common, we focus our attention on this technique, but point readers to **ref. [168]** to learn more about IPES/UPS. Versions of this technique have been used on a variety of materials such as TiS$_3$, [227] carbon nanotubes, [228] and highly oriented pyrolytic graphite (HOPG). [229] For this section we are primarily concerned with STS measurements which have yielded trustworthy $E_g$ (and by extension $E_b$) values for organic semiconductors and TMDC monolayers.

STS uses a scanning tunneling microscope (STM) to probe the local density of states as electrons or holes are injected into a material. A metallic tip with an apex radius on the order of a few nanometers is brought into close contact with the sample surface and the voltage of the tip is varied while the tip height ($z$) and current ($I$) are monitored. In most cases, $z$ is held constant and the derivative of the $I$ vs. $V$ spectrum is related to the density of states in the sample, which is then used to determine $E_g$. [230] However, oftentimes for organic semiconductors it is advantageous to hold $I$ constant while varying $z$, which produces a $z$-$V$ spectrum that can be used in a similar manner to measure $E_g$. [172] The challenge in each case boils down to interpreting the STS measurement; that is to say, given an STS signal, how does one properly interpret it to obtain the CB and VB energies and hence the band gap energy. In order to provide examples of this process, we highlight two STS studies from the literature. The first is Alvarado et al's measurement of $E_b$ on organic semiconductors (PFO and poly(1,12)AOPV-co-PPV)[166] using $z$-$V$ scans, and the second is Rigosi et al's $E_b$ measurement on TMDC monolayers using $I$-$V$ scans.[87] Both of these studies were sorely needed at their time of publication, 1998 and 2015 respectively, due to the fact that communities were split over conflicting $E_b$ values (measurements ranged from ~200 – 900 meV at their time of publication). Despite the low spectral resolution of STS and the corresponding high uncertainty in $E_b$ (~100 meV), these STS studies provided timely, trustworthy estimates of $E_b$ when optical spectroscopies either failed to measure $E_b$ or needed additional verification.

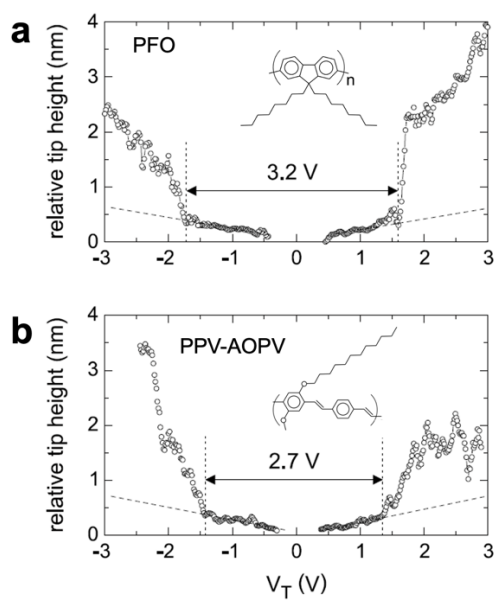

**Figure 20**: Typical tip height vs voltage curves for two organic semiconductors: **(a)** PFO and **(b)** poly(1,12)AOPV-co-PPV. Data was collected in constant current mode with tunneling currents held constant at either 50 or 100 pA. The abrupt steps in tip height at negative (positive) voltages indicate tunneling into the semiconductor, and formation of electron (hole) polarons. The difference between electron and hole polaron energies is defined to be the band gap. Panels (a-b) are reprinted from ref. [166] with permission from the American Physical Society, Copyright 1998.

In the 1998 study by Alvarado et al.,[166] very thin films (2 – 12 nm) of conjugated polymers PFO and PPV-AOPV were created by spin-coating onto a Au(111) surface (see **Figure 20** for the molecular structure and **Table 4** for the chemical formula).[166] Typical tip height vs voltage curves obtained for the two materials are shown in **Figure 20**. At small negative tip biases, electrons from the Fermi level of the STM tip can only access vacant states above the Fermi level of the gold substrate. However, as the tip bias is made more negative, electrons can eventually tunnel into vacant states in the polymer; thus, the abrupt change in tip height at a particular negative voltage indicates the formation of the electron polaron, analogous to an electron tunneling into the conduction band of a conventional semiconductor. Similar arguments apply for positive polarity and the hole polaron, and the difference between the two polaron states is defined to be the transport gap which is indicated by the double-headed arrows.

A series of independent measurements was made at different locations, and the estimated uncertainty based on local variations was ±100 meV for each of the two materials. The exciton resonance energy was determined from a broad, excitonic absorption peak in the room-temperature absorption spectrum, and this resulted in $E_b = 300 \pm 100$ meV for PFO and $E_b = 360 \pm 100$ meV for PPV-AOPV. These values are smaller than $E_b$ as measured by electroabsorption in conjugated polymers, and this is expected since STS measures the transport gap rather than fundamental gap, and the former is smaller due to the relaxation of free electrons and holes into polarons. [163]

Next we turn to the more recent work by Rigosi et al. from 2016, [87] in which five TMDC compounds were studied: $Mo_xW_{1-x}S_2$ for x = 0, 0.1, 0.25, 0.5, and 1. The TMDC samples were monolayers mechanically exfoliated onto fused quartz substrates from bulk crystals obtained from a commercial supplier. The transparent fused quartz was used to facilitate optical measurements of $E_{1s}$, but in order to also do STS the authors made an electrical connection to the samples by evaporating a ~20 nm thick gold contact through a TEM grid serving as a shadow mask (avoiding photoresist and solvents).

Typical $dI/dV$ curves obtained for the five materials are shown in **Figure 21a**. As with the previous example, for small negative tip biases electrons from the Fermi level of the STM tip cannot tunnel into the semiconductor, yielding small currents and $dI/dV$ values below what is shown. However, as the tip bias is made more negative, electrons can eventually tunnel into the material. Thus, the increase in $dI/dV$ as voltages get more negative indicates electrons tunneling into the CB. Similarly, for large enough positive voltages, holes can begin tunneling into the VB. The arrows indicate the CB and VB edges. However, in this case the band edges are not readily apparent from the data itself and a tunneling model is employed to fit the data and identify the edges. The tunneling model relies on knowing (or estimating) the electron and hole effective masses, the work functions of the semiconductor and STM tip, the tip-to-sample distance, the crystal momentum at the K/K′ points of the TMDC band structure, and an assumption that the tunneling is dominated by the behavior at those points. The band edges are the adjustable parameters obtained from doing least squares fits of their model to their data.

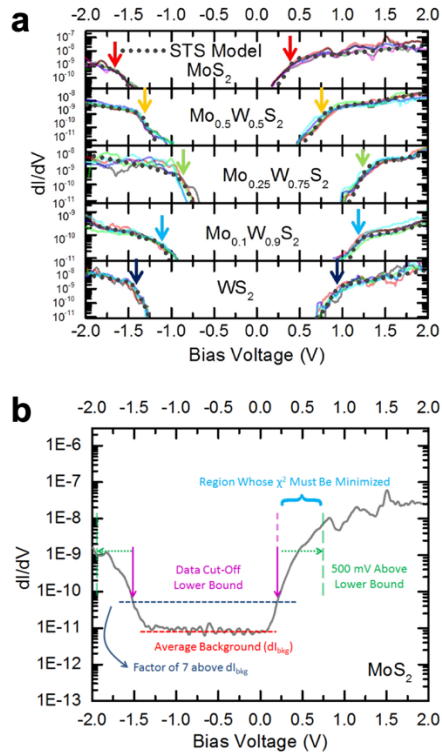

**Figure 21**: **(a)** Typical $dI/dV$ vs voltage curves for five TMDC samples. Data was collected by turning off the feedback for the tip height and recording the tunneling current while quickly ramping the tunneling voltage. The solid lines are averages of data measured at different randomly selected spatial locations; the dotted lines are fits to a theoretical model. The arrows indicate the CB and VB edges as established by the model; the difference between the two (minus an additional tip-induced band bending correction) is the band gap energy. **(b)** Illustration of how the portion of the STS spectra is selected for the fit, using representative data for $MoS_2$. The fitting range goes from the voltage which provides a signal 7× above the background, to 0.5 V above that. Panels (a-b) are reprinted from ref. [87] with permission from the American Physical Society, Copyright 2016.

In order to do the fits, a fitting range within the STS spectrum must be selected, and **Figure 21b** illustrates how the authors did this. Specifically, the dashed red line shows the background signal, i.e. the tunneling which occurs within the band gap. The starting point for the fit is given by the voltage where the value of $dI/dV$ has increased by 7× above the background. The ending point for the fit is 0.5 V away from that. These two limits were chosen so as to avoid undue influence from noise and defect states (the factor of 7) and to avoid tunneling from parts of the Brillouin zone away from the band edges (limiting to a 0.5 V range); however, these specific numerical factors were somewhat arbitrarily chosen based on experience and reasonableness of the fit. The authors also take into account the tip-induced band bending, which is calculated (via a simulation tool) to be ~30 meV; that value is subtracted from the band gap inferred from the band edges fit from the STS spectra. The uncertainties in their deduced band gaps are reported to be between ±40 and ±60 meV, although it is unclear if that is only the uncertainty arising from the standard deviation from the measurements made at different locations, or if they also include uncertainties arising from the fit procedure.

To summarize, STS can be used to obtain trustworthy values of $E_g$ (and by extension $E_b$) when optical detection of $E_g$ is not possible. Due to the experimental challenges associated with STS and its poor spectral resolution, it makes sense to only resort to STS when optical methods fail or need additional verification. However, if the STS data is properly acquired and IS modeled correctly, then this technique is capable of yielding trustworthy measurements of the exciton binding energy.

## 9. Challenges with identifying the band gap through optical absorption

Although the relationship between $E_b$ and the optical absorption has already been introduced in **Section 2** and discussed intermittently throughout **Sections 5**, **6**, and **7**, we return for a moment to briefly summarize the challenges associated with identifying the band gap in the optical absorption, since this is often done incorrectly in the literature.

Assuming the Rydberg states are not resolved, it is very challenging to confidently assign $E_g$ (and by extension $E_b$) based solely on a semiconductor's optical absorption. To do so, an interband feature must be clearly resolved with a functional form predicted by Elliott's formula, e.g. a Heaviside step function convoluted with a broadening function.[17] Upon compiling this review article, we are only aware of one class of materials, 2D HPs, with an interband absorption feature that satisfies this condition. Moreover, even for 2D HPs the clarity of the interband feature is highly variable and only results in a confident assignment of $E_g$ and $E_b$ when the feature's origin is independently verified using other techniques.

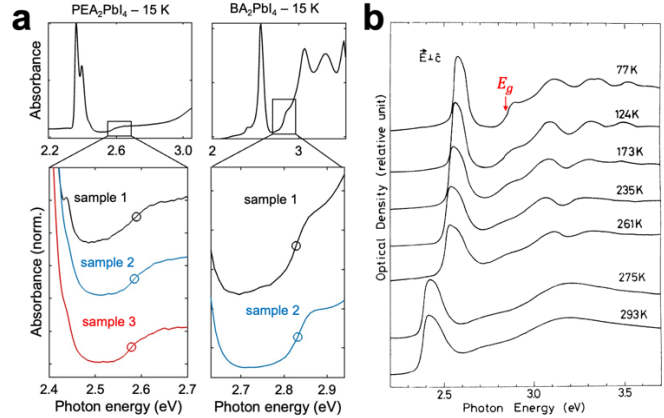

**Figure 22: (a)** A zoom-in on the interband absorption of several 2D HP thin films. The open circle marks $E_g$. **(b)** Optical absorption at various temperatures of a similar 2D HP with composition $(C_{10})_2PbI_4$ where $C_{10}$ = Decylammonium. The same interband absorption featured in panel (a) becomes clear at low temperature. Adapted from ref. [121] with permission from the American Physical Society, Copyright 1990.

Several examples of the 2D HP interband absorption feature are shown in **Figure 22a** for 2D HP thin films with $BA_2PbI_4$ and $PEA_2PbI_4$ compositions. These spectra are from a 2023 electroabsorption study by Hansen et al. [5] and the open circles mark independent $E_g$ assignments that were obtained via the EA-based method highlighted in **Figure 15**. Due to the two-dimensional nature of these semiconductors, the interband absorption is expected to (and does) resemble a Heaviside step function convolved with a broadening function, where the mid-point of the step corresponds to $E_g$. [17] And the confirmation from EA that $E_g$ indeed lies at the step's midpoint renders the feature's origin unambiguous. However, as unambiguous as this assignment now seems in hindsight, prior to this confirmation from EA, versions of this same feature in various 2D HPs were sometimes correctly assigned to $E_g$ [19, 121, 231] and other times incorrectly assigned to the $2s$ exciton state. [104, 107, 117] The absorption spectrum at 77 K in **Figure 22b** is an example of one instance where $E_g$ was correctly assigned to the step-like feature in a 1990 study by Ishihara et al. [121] However, without confirmation from EA, the feature's origin is not obvious and the authors also independently measured $E_b$ via temperature dependent PL to corroborate their assignment of $E_g$ in the absorption.

More commonly, the optical absorption does not contain sufficient information to assign $E_b$. Even for 2D HPs, the interband feature highlighted in **Figure 22** only appears for some compositions and only when the sample is of high quality and cooled to low temperature. In cases where the interband feature isn't resolved, we find that oftentimes studies still attempt to assign $E_g$ and $E_b$ based on insufficient information, resulting in poor $E_b$ measurements. The lack of recognition of the uncertainties inherent to these optical absorption measurements may stem from confusion surrounding the required precision of $E_g$ in the equation $E_b = E_g - E_{1s}$. For example, to measure $E_b$ with 10% uncertainty via independent measurements of $E_{1s}$ and $E_g$, one must determine $E_g$ very precisely with an uncertainty of $E_b/10$ or better. This required precision is often overlooked because $E_g$ itself is most often measured by the broad absorption onset with a large absolute uncertainty ($\sigma_{Eg} \sim 100$ meV), but a low relative uncertainty ($\sigma_{Eg}/E_g \sim 0.05$). Such measurements are perfectly legitimate estimates of $E_g$, but are problematic when applied to the equation $E_b = E_g - E_{1s}$ since $\sigma_{Eg}$ and $E_b$ are often of the same order-of-magnitude, with $\sigma_{Eg}$ being larger than $E_b$ in many cases. In the following, we highlight three recent $E_b$ measurements where $\sigma_{Eg} \gtrsim E_b$, but the studies failed to notice this and did not report an uncertainty for $E_g$ or $E_b$.

In a 2022 study by Niedzwiedzki et al.,[6] multiple techniques were used to measure $E_b$ on the same MAPbI$_3$ thin film to studythe variation in $E_b$ between different measurement methods. Fitting to Elliott's equation was among the techniques, and their fit is displayed in **Figure 23a** for a MAPbI$_3$ thin film at 78 K. The data (solid black) and fit (dashed red) lines are in reasonable agreement. The authors report $E_b = 17$ meV but do not report an uncertainty. We performed an Elliott fit to the same data independently and obtained $E_b = 12 \pm 44$ meV; i.e., the uncertainty in $E_b$ is greater than $E_b$ itself. About a dozen other studies have measured $E_b$ on MAPbI$_3$ in a similar manner (**Table 3**); however, each suffers from the same problem of broad spectral features, uncertainty in the position and widths of the Rydberg members, and uncertainty in the magnitude of the Urbach tail. Thus, even at low temperatures the absorption of MAPbI$_3$ is insufficient to accurately determine $E_b$ and this explains the large variance in reported $E_b$ among studies that use absorption fitting (**Table 3**).

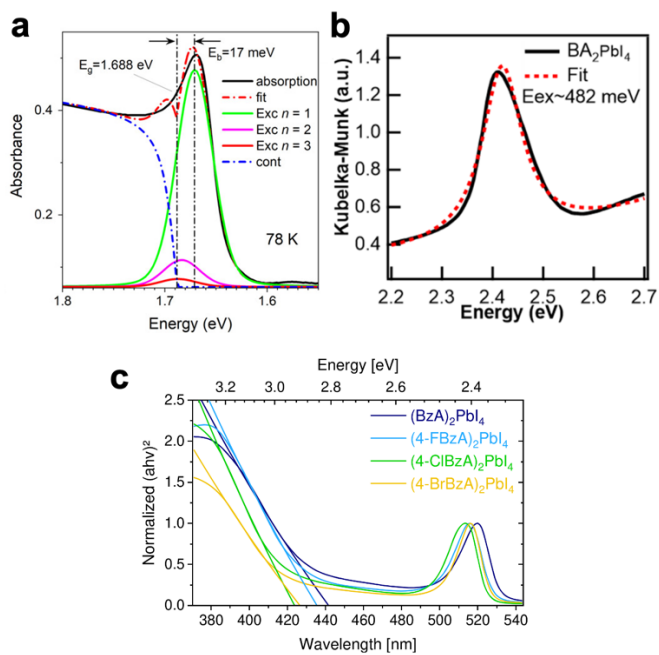

**Figure 23: (a)** Optical absorption of a MAPbI$_3$ thin film at 78 K, fit to Elliott's equation. Reprinted from ref. [6] with permission from the American Chemical Society, Copyright 2022. **(b)** Optical absorption of a 2D HP powder also fit to Elliott's equation. Reprinted from ref. [154] with permission from the American Chemical Society, Copyright 2021. **(c)** Tauc plot of $(AE)^2$ where $A$ is the absorbance and $E$ is the photon energy. A line is fit to the broad, above-gap absorption and $E_g$ is extracted from the x-axis intercept. This Tauc method of estimating $E_g$ should not be used in $E_b$ measurements since the uncertainty in $E_g$ is greater than $E_b$ itself. Reprinted from ref. [232] with permission from the American Chemical Society, Copyright 2020.

This same type of fit procedure has also been applied to 2D HPs in a similarly misguided way, as shown in **Figure 23b**. Unlike the BA$_2$PbI$_4$ thin film absorption in **Figure 22a**, the absorption of the BA$_2$PbI$_4$ powder in **Figure 23b** (from Ref. [154]) does not exhibit a clear interband feature. The authors performed an Elliott fit to extract $E_g$, however, this results in a value of $E_b$ (482 meV) that is inconsistent with clear EA measurements on the same material (256 meV). [5] Unfortunately, it is also somewhat common practice for $E_g$ to be determined using the Tauc method and then used to determine $E_b$, e.g. as shown in **Figure 23c** for a handful of lead-iodide 2D HPs. In this 2020 study, [232] the authors report $E_b$ values of ~475 meV for each of the compounds, without giving uncertainty estimates, and these values are inconsistent with clear EA measurements on lead-iodide-based 2D HPs. [5] Similar to the Elliott fit, this method yields only an approximate measurement of $E_g$ that should not be used to determine $E_b$. Unlike the Elliott fit, however, this method is also conceptually flawed because the functional forms used in Tauc fits only apply to non-excitonic cases. Moreover, the choice of the fit range in the above-gap absorption is somewhat arbitrary and the slopes of the fitted feature can originate from multi-valley absorption which is unrelated to the location of $E_g$.

**Conclusions**

We have evaluated and compared the measurement quality of six techniques for measuring $E_b$: absorption fitting, temperature dependent PL, resolving Rydberg states, EA spectroscopy, magneto-optical spectroscopy, STS, and identifying the interband transition.

We have shown that temperature-dependent PL is not an ideal method for measuring $E_b$, as there are many nested assumptions in the converting the Arrhenius activation energy to $E_b$ and the technique yields very inaccurate $E_b$ values for 2D HPs. While temperature-dependent PL may in some instances be a viable method for measuring $E_b$, the degree to which the assumptions are satisfied for a given material is nearly always shrouded in uncertainty. The remaining techniques: resolving Rydberg states, EA spectroscopy, magneto-optical spectroscopy, and scanning tunneling spectroscopy are all capable of producing extremely trustworthy $E_b$ measurements provided the experiments are correctly and carefully conducted. In **Figure 24a**, we have plotted the approximate relative accuracy and precision of the various measurement techniques according to our best judgment based on reading 100+ studies from the literature that report $E_b$ measurement. As discussed previously, resolving Rydberg states is an effective method when the origin of band-edge spectral features is clear. The accuracy of such measurements, however, depends entirely on the confidence of the assignments— if the Rydberg features are reproducible sample-to-sample and/or the features exhibit a magnetic or electric-field dependence expected for an exciton then the measurement can be considered very accurate. The precision, on the other hand, is not as good as EA or magneto-optical spectroscopies or due to the fact that a model must be assumed to extrapolate the Rydberg energy ladder to $E_g$ in order to obtain $E_b$. When Franz-Keldysh features are resolved via EA spectroscopy (in the $E_b/\Gamma > 2$ regime) or Landau-level transitions are resolved via magneto-optical spectroscopy (in the high-field regime), then $E_g$ and $E_b$ can be measured accurately and precisely in an unambiguous, model-independent fashion. When optical techniques fail to produce clear measurements, as is often the case, STS is an excellent option to provide a rough measurement of $E_b$ to within ~50-100 meV.

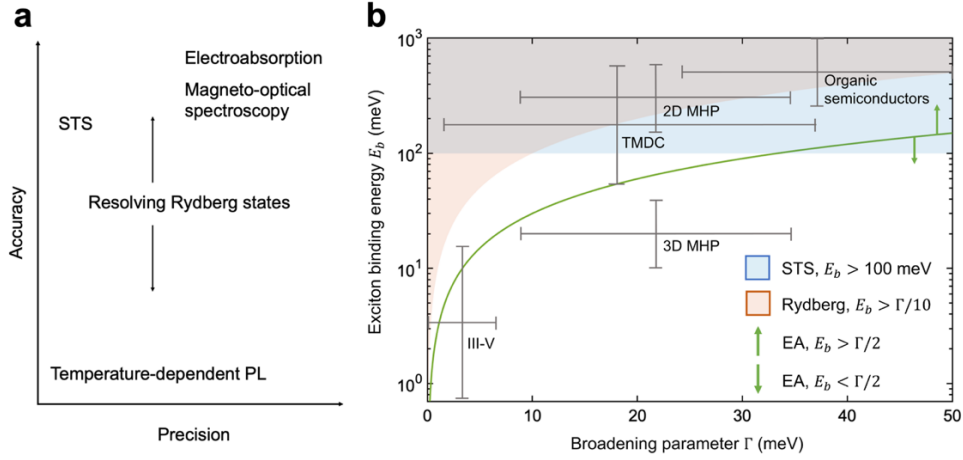

**Figure 24**: **(a)** Approximate ranking of the $E_b$ measurement techniques by accuracy and precision. The data point for electroabsorption assumes the $E_b/\Gamma > 2$ regime where Stark and Franz Keldysh features are individually resolved. The datapoint for magneto-optical spectroscopy assumes the high-field regime where Landau level-associated transitions are resolved. **(b)** The parameter space of $E_b$ vs $\Gamma$ with approximate bounds shown for the classes of semiconductors discussed in this article.

Several factors must be considered when selecting the appropriate measurement technique for a given material. In **Figure 24b**, the parameter space of $E_b$ vs $\Gamma$ is plotted and the approximate range occupied by various classes of semiconductors is indicated by the gray lines. Due to the large experimental error in STS measurements, this technique should only be considered when $E_b$ is large ($\gtrsim$ 100 meV). Resolving Rydberg states, on the other hand, can be done for any range of $E_b$ provided the broadening $\Gamma$ is sufficiently low. If $E_b/\Gamma > 10$, then this technique is quite likely to be successful. EA spectroscopy is also sensitive to the ratio of $E_b/\Gamma$, with the best measurements coming in the $E_b/\Gamma > 2$ regime. The other EA regime ($E_b/\Gamma < 2$) has produced excellent measurements for MAPbI$_3$ and GaAs, but is relatively new and untested in its application to other materials. Magneto-optical methods are also somewhat sensitive to $E_b/\Gamma$, but depend much more on $m^*$ and therefore have been excluded from the plot.

By considering $E_b$ measurements from "material property" and "experimental technique" perspectives, a picture emerges that will help guide $E_b$ measurements in the future. Furthermore, the evaluation of the experimental techniques contained herein also serves the practical purpose of assessing $E_b$ measurements for TMDC monolayers and HP. After a decade of significant research efforts focused on these materials, it is clear that $E_b$ ~350 - 450 meV for TMDC monolayers between SiO$_2$ and vacuum, $E_b$ ~ 150 - 200 meV for TMDC monolayers encapsulated in hBN, $E_b$ ~ 200 - 300 meV for lead-iodide 2D HP compositions such as BA$_2$PbI$_4$ and PEA$_2$PbI$_4$, and $E_b \sim 10$ meV for MAPbI$_3$. Confidence in these values has not come easy, as widely conflicting measurements continue to persist in the literature. However, when the data and measurement quality are considered en masse, it is clear that these values appear most often and also appear in the experiments with the clearest signals.

**Conflicts of interest**

There are no conflicts to declare.

**Acknowledgements**

KRH acknowledges support from the National Science Foundation through a Graduate Research Fellowship (Grant No. 1747505). LWB acknowledges the Sloan Foundation through an Alfred P. Sloan Research.